\documentclass[aps,prl,reprint,superscriptaddress]{revtex4-1} 

\usepackage{amsmath,amssymb}
\usepackage{graphicx}
\usepackage{dcolumn}
\usepackage{bm}

\usepackage[utf8]{inputenc}
\usepackage[T1]{fontenc}
\usepackage{mathptmx}
\usepackage{epstopdf,xcolor}
\usepackage{appendix}

\begin{document}


 \title{Universal Analytic Model of Irradiation Defect Dynamics in Silica-Silicon Structures}

\author{Yu Song}
\email[Corresponding author: ]{yusong@njtc.edu.cn} 
\affiliation{College of Physics and Electronic Information Engineering, Neijiang Normal University, Neijiang 641112, China}
\affiliation{Microsystem and Terahertz Research Center \& Institute of Electronic Engineering, China Academy of Engineering Physics, Chengdu 610200, China}

\author{Guanghui Zhang}
\affiliation{Microsystem and Terahertz Research Center \& Institute of Electronic Engineering, China Academy of Engineering Physics, Chengdu 610200, China}

\author{Xue-Fen Cai}
\affiliation{Beijing Computational Science Research Center, Beijing 100193, China}

\author{Yang Liu}
\affiliation{Microsystem and Terahertz Research Center \& Institute of Electronic Engineering, China Academy of Engineering Physics, Chengdu 610200, China}

\author{Hang Zhou}
\affiliation{Microsystem and Terahertz Research Center \& Institute of Electronic Engineering, China Academy of Engineering Physics, Chengdu 610200, China}

\author{Le Zhong}
\affiliation{Microsystem and Terahertz Research Center \& Institute of Electronic Engineering, China Academy of Engineering Physics, Chengdu 610200, China}

\author{Gang Dai}
\affiliation{Microsystem and Terahertz Research Center \& Institute of Electronic Engineering, China Academy of Engineering Physics, Chengdu 610200, China}

\author{Xu Zuo}
\affiliation{College of Electronic Information and Optical Engineering,
Nankai University, Tianjin 300071, China}

\author{Su-Huai Wei}
\email[Corresponding author: ]{suhuaiwei@csrc.ac.cn}
\affiliation{Beijing Computational Science Research Center, Beijing 100193, China}

\date{\today}

\begin{abstract}
Irradiation damage is a key physics issue for semiconductor devices under extreme environments.
For decades, the ionization-irradiation-induced damage in transistors with silica-silicon structures under constant dose rate is modeled by a uniform generation of $E'$ centers in the bulk silica region and their irreversible conversion to $P_b$ centers at the silica-silicon interface. But, the traditional model fails to explain experimentally observed dependence of the defect concentrations on dose, especially at low dose rate. Here, we propose that, the generation of $E'$ is decelerated due to the dispersive diffusion of induced holes in the disordered silica and the conversion of $P_b$ is reversible due to recombination-enhanced defect reactions under irradiation. It is shown that the derived analytic model based on these new understandings can consistently explain the fundamental but puzzling dependence of the defect concentrations on dose and dose rate in a wide range.
\end{abstract}

\maketitle

The electrical properties of the most-widely-used silicon devices degrade in the outer space and other extreme environments~\cite{adell2014dose}.
This is because persistent ionizing irradiation induces
$E’$ centers in the dielectric silica layer and $P_b$ centers at the silica-silicon interface, respectively~\cite{lenahan1984hole,poindexter1984electronic}.
The dominating $E’$ centers are positively charged states of oxygen vacancies with puckered configuration ($V_{O\gamma}$)~\cite{weeks1956paramagnetic,lenahan1984hole},
while the $P_b$ centers are positively charged silicon dangling bonds~\cite{poindexter1981interface,poindexter1984electronic}.
These defects act as oxide and interface traps in the silica-silicon structures~\cite{fleetwood1993effects}, which alter the performance of the devices.
To understand the microscopic picture and model the irradiation-induced defect dynamics is essential to evaluate, predict, and control the irradiation damage.
Since the pioneer work of Mclean in 1980~\cite{mclean1980framework}, a series of works have made earnest efforts to determine the atomic-scale physical mechanisms.
For reviews, see Refs.~\cite{oldham2003total,fleetwood2017evolution,adell2014dose}.

It has been proposed that, $E'$ and $P_b$ centers are produced by
four main processes in the silica-silicon structures~\cite{oldham2003total}.
Firstly, the irradiation generates electron-hole-pairs in the silica region, and a fraction of which will subject to prompt
recombination~\cite{Shaneyfelt1991charge,boch2006temperature}.
Secondly, the remaining holes transport via hopping among relatively shallow defects such as oxygen vacancies with dimer configuration ($V_{O\delta}$) broadly distributed in silica~\cite{hughes1977time,boesch1975hole,nicklaw2002structure}
and are captured by relatively deep defects such as $V_{O\gamma}$ concentrated near the interface~\cite{lenahan1984hole,conley1993room,conley1994observation_JAP,
blochl2000first,lu2002structure}.
$V_{O\delta}^+$ has a large electron capture cross section and can act as recombination centers~\cite{conley1994observation,conley1994observation_JAP,yang2016non}.
Thirdly, the generated $E’$ centers  ($V_{O\gamma}^+$) can
crack molecule hydrogen ($\rm H_2$) in the system and release protons ($\rm H^+$)~\cite{conley1993room,conley1993molecular,stahlbush1993post,
tuttle2010defect}.
Fourthly, the released protons travel to the interface where they de-passivate H-passivated dangling bonds ($P_bH$) by forming $P_b$ centers and $\rm H_2$~\cite{stahlbush1993post,mrstik1991si,stathis1994atomic,rashkeev2001defect}.
These processes formed the bases for the modeling and simulation of the irradiation damage~\cite{hjalmarson2003mechanisms,
rowsey2011quantitative,rowsey2012radiation,
rowsey2012mechanisms,hughart2012effects}.
These processes can be described as a generation-conversion framework:
$h^+\stackrel{k_0}{\rightarrow} E' \stackrel{k_{f1}}{\underset{k_{b1}}{\rightleftarrows}} H^+ \stackrel{k_{f2}}{\rightarrow} P_b$,
where $k_{f,b}$'s denote forward (f) and backward (b) conversion rate constants~\cite{fahey1989point}.
$k_0=2D_{h0}L_c$ is assumed to be a \emph{uniform} generation rate constant, where $D_{h0}$ is the diffusion coefficient of holes in silica and $L_c$ is an estimated critical length similar to the concept of capture cross section.
The conversion from $E'$ to $\rm H^+$ 
is reversible due to small reaction barriers for both directions~\cite{tuttle2010defect}. 
However, the conversion from $\rm H^+$ to $P_b$ is assumed to be irreversible as the backward reaction barrier is considered to be large (about 1.3eV)~\cite{rashkeev2001defect}. Accordingly, the whole backward conversion from $P_b$ to $E'$ is assumed to be negligible.

This uniform-generation and irreversible-conversion model has dominated the field for more than 3 decades.
However, we notice that the fundamental dependence of the concentration of $E'$, [$E'$], on the irradiation dose cannot be properly explained by this ``standard'' picture.
A monotonous growth profile is expected for [$E'$] according to the traditional model, 
however, an initial increase and then decrease behavior is recently found at low dose rate (at 10 mrad/s)~\cite{li2019correlation}.
This abnormal non-monotonous dose dependence cannot be explained by the traditional model, unless extra nonradiative recombination of charge carriers, which is normally expected for high-dose-rate case, is assumed \emph{exclusively} for the low-dose-rate case~\cite{li2019correlation}.
Moreover, a striking negative dose-rate dependence (or enhanced low-dose-rate sensitivity, ELDRS) has been experimentally found for irradiation-induced [$P_b$] in 1991~\cite{Enlow1991Response},
which becomes the main obstacle for people to deduce the extremely long time low-dose-rate irradiation damage from the high-dose-rate experiments carried out at relatively short time.
Since then, great efforts have been made and many possible mechanisms have been proposed~\cite{fleetwood1994physical,rashkeev2002physical,
hjalmarson2003mechanisms,rowsey2012mechanisms,
boch2006dose}.
However,till now, there is still no solid experimental evidence supporting these claimed mechanisms.

In this Letter, we introduce two new concepts, \emph{decelerated generation of $E'$} and \emph{reversible conversion from $E'$ to $P_b$}, in the framework of irradiation defect dynamics,
\begin{equation}\label{eq:generation}
{\rm h^+} \stackrel{k(t)}{{\rightarrow}} E'
\stackrel{k_{f1}(q)}{\underset{k_{b1}(q)}{\rightleftarrows}} {\rm H^+}
\stackrel{k_{f2}(q)}{\underset{k_{b2}(q)}{\rightleftarrows}} P_b,
\end{equation}
where a time-dependent generation rate constant $k(t)$ and dose-rate ($q$)-dependent conversion rate constants $k_{f,b}(q)$
are introduced that are distinct from the existing models.
The physical bases are the dispersive diffusion of irradiation-induced holes in the disordered silica and recombination-enhanced conversion reactions under irradiation, respectively.
Based on these new concepts, we derive a new analytic model of the dose dependence of the defect concentrations and test it through $\gamma$-ray irradiation experiments on gated lateral PNP transistors.
A perfect match is found for an extremely wide dose rate range from 58 $\mu$rad/s to 1 rad/s, which not only verify our new concepts but also provides insights into the origin of the dose rate dependence.

The concept of decelerated generation of $E’$ centers is proposed based on the well-known dispersive feature of the hopping transport of holes in disordered silica~\cite{hughes1977time,boesch1975hole,nicklaw2002structure}.
In disordered silica the difference in atomic configurations of localized $V_{O\delta}$ states results in an exponential distribution of activation energy for holes~\cite{bendler1985derivation}.
In this situation, the holes undergo a continuous-time random walk (CTRW) among localized states, 
and the pausing-time has a broad distribution of $\psi(t)\propto t^{-(1+\alpha)}$, where $0<\alpha<1$ is a dispersion parameter~\cite{scher1975anomalous,shlesinger1984williams}.
As a result of CTRW, the diffusion coefficient of the holes is not a constant but a power-law decay function of time, $D_h(t)=D_{h0}(w_0t)^{-\alpha}$, where $w_0$ is an attempt frequency.
This unique behavior has been experimentally observed~\cite{pfister1978dispersive,scher1975anomalous,kakalios1987stretched}.
From the generation reaction of $h^+ + V_{O\gamma} \rightarrow E'$, the rate equation of [$V_{O\gamma}$] is given by $ d[V_{O\gamma}](t)/dt=-k(t) [h] [V_{O\gamma}](t)$, where $k(t)=k_0(w_0t)^{-\alpha}$
and $[h]$ is the equilibrium concentration of holes under irradiation and recombination~\cite{fowler1956x,chen1981solution}.
The energy levels of the recombination centers such as $V_{O\delta}$ distribute exponentially~\cite{nicklaw2002structure}, so
$[h]=c q^{0.5}$~\cite{fowler1956x,fowler1954conductivity-1}.
The time dependence of [$V_{O\gamma}$] is solved as a stretched exponentially decay function~\cite{williams1970non,kohlrausch1854theorie},
[$V_{O\gamma}$]$(t)=P_0e^{-k’ \sqrt{q}t^\beta}$, where $P_0$ is the initial concentration of $V_{O\gamma}$,
$\beta=1-\alpha$, and  $k’=c\beta^{-1}w_0^{-\alpha} k_0$.
The concentration of $E'$ is then given by $[E’](t)=P_0-[V_{O\gamma}](t)$, 
which reduces to a fractional power law (FPL) dose dependence, $[E’]=k’P_0\sqrt{q}t^\beta$, for a relatively short time.
Such a general FPL dependence with $\beta$<1 implies a decelerated generation of $E’$ centers and has been observed in bulk silica~\cite{griscom1990growth,imai1994intrinsic,mashkov1996fundamental}.
For comparison, in the case of constant diffusion coefficient 
as suggested in traditional models, $[P]$ is a simple exponential decay function of time,
$[P]=P_0e^{-k_0c\sqrt{q}t}$, and $[E’]$ displays a linear dependence on time, $[E’]=k_0P_0 c\sqrt{q}t$.
Thus, the concept of decelerated generation of $E’$ follows directly from the dispersive diffusion mechanism.
The generation equation as a function of dose ($D=qt$) reads
\begin{subequations}
\begin{equation}
(d E'/d D)_g = g_e \beta D^{\beta-1},
\end{equation}
where $g_e= k'P_0q^{0.5-\beta}$ is an effective generation efficiency.
This result also implies a dose rate dependence of $q^{0.5-\beta}$ for a fixed $D$, which is $q^{-0.5}$ in the case of traditional model.

The concept of a reversible conversion between $E'$ and $P_b$ centers is based on the fact that the conversion reactions depend on the irradiation. Under the irradiation, the non-radiative recombination of induced carriers at the broadly distributed $V_{O\delta}$~\cite{conley1994observation,conley1994observation_JAP} can release energy and enhance all the four conversion rate constants 
and promote a nonzero $k_{b2}(q)$ in Eq. (1)~\cite{kimerling1978recombination,weeks1975theory}.
This so-called recombination-enhanced defect reaction mechanism has been observed in many experiments~\cite{lang1974observation,zhang1995multiple}.
The re-generated protons can go back to the near-interface region and convert to $E'$ with a rate constant of $k_{b1}(q)$, as indicated in Eq. (1).
Hence, the whole conversion between $E'$ and $P_b$ becomes reversible.
From the conversion reactions of 
$E' + H_2 \rightleftarrows EH + H^+$ and $H^+ + P_bH \rightleftarrows H_2 + P_b$, the rate equations can be obtained as
$(d E'/d t)_c= - k_{f1}{\rm [H_2]} E' + k_{b1}{\rm [EH]} {\rm [H^+]}$,
$d {\rm [H^+]}/d t=k_{f1}{\rm [H_2]} E' - k_{b1}{\rm [EH]} {\rm [H^+]} -k_{f2}{\rm [P_bH]} {\rm [H^+]} + k_{b2}{\rm [H_2]} P_b$, and
$d P_b/d t= k_{f2}{\rm [P_bH]} {\rm [H^+]} - k_{b2}{\rm [H_2]} P_b$.
Here $\rm H_2$ can be recycled and the initial [$\rm EH$] 
and [$P_bH$] are much higher than the changes due to the reactions~\cite{rowsey2011quantitative}, thus they keep almost constant during the irradiation process.
Note that
the protons are simultaneously released and annihilated,
a \emph{quasi-steady-state} condition can be assumed for protons: $d {\rm [H^+]}/d t=0$.
Then the rate equations of $E'$ and $P_b$ can be reduced as
\begin{equation}\label{eq:interconcersion}
(d E'/d D)_c = -d P_b/d D = -q^{-1}\kappa_f E' + q^{-1}\kappa_b P_b, 
\end{equation}
\end{subequations}
where $\kappa_f = k_{f1}{\rm [H_2]} k_{f2}{\rm [P_bH]}/(k_{b1}{\rm [EH]} + k_{f2}{\rm [P_bH]}) $ and $\kappa_b = k_{b2}{\rm [H_2]}k_{b1}{\rm [EH]} /( k_{b1}{\rm [EH]} + k_{f2}{\rm [P_bH]}) $ are effective conversion rate constants.

The solution of Eq. (2b) with 
$E'(0)=E_0$ and $P_b(0)=0$ provides insights in the dose and dose rate dependence of the conversion process. 
The results are $E’(D)=(1-\lambda)E_0+\lambda E_0 e^{-D/D_c}$ and $P_b(D)=\lambda E_0-\lambda E_0 e^{-D/D_c}$, where  $\lambda=\kappa_f/(\kappa_f+\kappa_b)$ and $D_c = q/(\kappa_f+\kappa_b) $.
From these results, we can see that the ratio between the irradiation time $t_r=D/q$ and the characteristic conversion time $\tau_c=1/(\kappa_f+\kappa_b)$ 
is $D/D_c$. So, $D_c $ is a characteristic conversion dose.
For $t_r\gg \tau_c$ ($D\gg D_c$), the conversion process is complete, and [$P_b$] is balanced at $P_b=\lambda E_0$. So, $\lambda$ is a conversion ratio from $E'$ to $P_b$.
As the dose rate increases, all the four $k_{f,b}$'s 
should increase because the non-radiative recombination increases.
However, the forward conversion reaction barriers in Eq. (1) are lower than the backward conversion reaction barriers~\cite{tuttle2010defect,rashkeev2001defect}, so, the effect of the non-radiative recombination energy on 
$\kappa_f$ should be less than on $\kappa_b$.
This will result in a decreasing $\lambda$ as the dose rate increases.
For comparison, in the traditional model, no dose rate dependence is present in the conversion ratio, i.e., $\kappa_b=0$ and $\lambda=1$ for any dose rate.

To verify these new concepts, 
we have derived an analytic model of the defect concentrations as a function of the irradiation dose.
This can be done by combining Eqs. (2a) and (2b) and using the initial conditions of $E'(0)=P_b(0)=0$~\cite{lenahan1984hole,poindexter1984electronic}.
The results are
\begin{subequations}
\begin{equation}
 E' (D) = (1-\lambda) g_e D^\beta+
\lambda g_e D_c^\beta e^{-D/D_c} \beta \Gamma[\beta,0,D/D_c],\\ 
\end{equation}
\begin{equation}
 P_b (D) = \lambda g_e D^\beta
- \lambda g_e D_c^\beta e^{-D/D_c} \beta \Gamma[\beta,0,D/D_c]. 
\end{equation}
\end{subequations}
This new analytic model of Eq. (3) contains only 4 effective parameters: $\beta$ and $g_e$ for defect generation, and $D_c$ and $\lambda$ for defect conversion.
$\Gamma(\beta,0, D/D_c) = \int_0^{D/D_c} x^{\beta-1}e^{-x}dx$ is a generalized incomplete gamma function in terms of $\beta$ and $D/D_c$, which reflects the \emph{interplay} between the decelerated generation and reversible conversion dynamics.

\begin{figure}[!t]
  \centering
  \includegraphics[width=\linewidth]{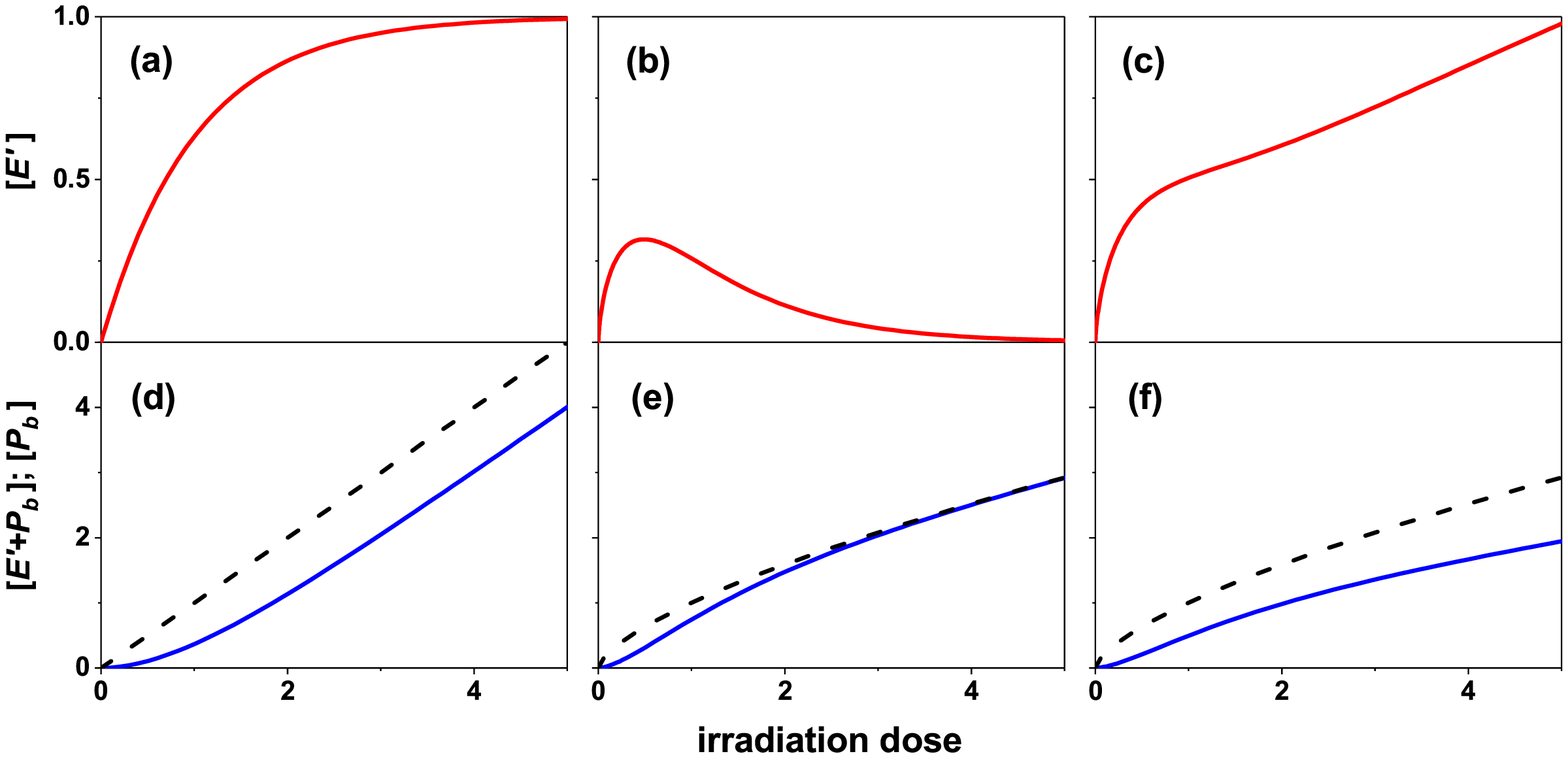}\\ 
  \caption{Dose dependence of [$E'+P_b$] (black), [$E'$] (red), and [$P_b$] (blue)
for (a, d) $\beta=\lambda=1$,
 (b, e) $\beta=2/3$, $\lambda=1$,
and (c, f) $\beta=\lambda=2/3$.
For all cases $g_e$ and $D_c$ are set as 1. } \label{ }
\end{figure}

If neither new concepts were introduced in Eq. (1), i.e., $\beta=\lambda=1$, the above dose dependence reduces to $E' (D) = g_e D_c (1-e^{-D/D_c})$ and $P_b (D)=  g_e D - g_e D_c (1-e^{-D/D_c})$ as plotted in Fig. 1 (a) and (d), respectively.
It is clear that, the total concentration (black dashed) increases linearly as the dose increases and [$E'$] (red solid) increases asymptotically as $E’$ centers are simultaneously generated and annihilated.
If only the concept of decelerated generation is introduced ($\beta<1$ and $\lambda=1$), the results become $E' (D) = g_e D_c^\beta \beta e^{-D/D_c} \Gamma[\beta,0,D/D_c]$ and $P_b (D) = g_e D^\beta - g_e D_c^\beta \beta e^{-D/D_c} \Gamma[\beta,0,D/D_c]$, as plotted in Fig. 1 (b) and (e), respectively.
It is seen that, the linear dependence of [$E'+P_b$] becomes an FPL dependence, and the asymptotical increase of [$E'$] becomes a non-monotonic dependent.
As the generation rate constant $k(t)$ decays with time, the continuous irreversible conversion makes [$E'$] decay for large dose.
If the concept of reversible conversion was further introduced ($\beta<1$ and $\lambda<1$), the results become Eq. (3) as plotted in Fig. 1 (c) and (f).
It is found that [$E'$] has an essentially different sublinear dependence on the dose, due to a sublinear FPL term from the backward conversion that dominates the dose profile.
It is also noticed that the near-linear dependence of [$P_b$] profile at big dose range
in Fig. (1d) (blue line) changes to a 
sublinear dependence in Figs. (1e) and (1f) due to $\beta<1$.

\begin{figure}[!t]
  \centering
  \includegraphics[width=\linewidth]{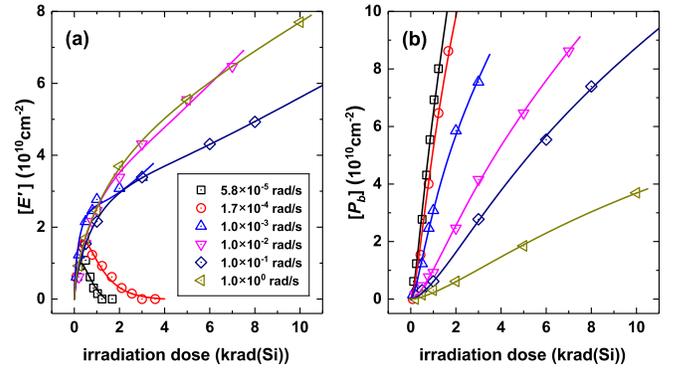}\\ 
  \caption{(a) [$E'$] and (b) [$P_b$] as a function of irradiation dose at different dose rate as indicated in the figure.
  Dots are for measured data and curves are for fitting.
} \label{ }
\end{figure}

To test the derived model, we measured [$E'$] and [$P_b$] by carrying out $\gamma$-ray irradiation experiments on lateral PNP transistors with gated silica-silicon structures~\cite{pease2004characterization} in an extremely wide range of dose rate.
The sum of [$E'$] and [$P_b$] is extracted from the voltage shift of sub-threshold sweep curves of the structure,  
while the value of [$E'$] can be deduced from the voltage shift of the peak position of the gate sweep curves~\cite{mcwhorter1986simple,ball2002separation}.  
The concentrations are investigated as a function of the irradiation dose and dose rate. $D$ is limited to 10 krad (Si) where the FPL applies. $q$ varies from 58 $\mu$rad/s to 1 rad/s, which spend 5 order of magnitudes. For each dose rate 3 samples are adopted to include sample-to-sample variability~\cite{song2020origin,song2020defect}.
The test at each dose is completed within half an hour, during which the annealing of $E'$ and $P_b$ can be ignored~\cite{adell2014dose}.

The typical data of [$E'+P_b$] are present in Fig. S1 in the supplementary materials (SM), from which a clear sublinear dose dependence is observed.
This behavior is totally different from the case of Fig. 1 (d), but is similar to the cases in Figs. 1 (e) and 1 (f), reflecting the necessity of including the concept of decelerated generation in the defect dynamics.
The data is fitted by the sum of Eqs. (3a) and (3b), $g_e D^\beta$, from which a perfect match is found in Fig. S1 in the SM.
The fitting parameters $\beta$ and $g_e$ are plotted in Fig. 3 (a) and (b) as a function of dose rate.
The parameter $\beta$ with an average value of about 2/3 is a direct result of the dispersive transport of holes ($\alpha=1/3$). 
The generation efficiency $g_e$ is found to decrease with a negative power law of the dose rate, $g_e \propto q^{-0.12}$. 
This result suggests a dose rate dependence of the hole concentration, $[h]\propto q^{0.55}$, which is consistent with the recombination on exponentially distributed defects~\cite{fowler1956x,fowler1954conductivity-1}.

The typical measured data of [$E'$] and [$P_b$] are shown in Fig. 2 (a) and 2 (b) as the dots.
The non-monotonous dependence of [$E’$] at relatively low dose rate is totally different from the curve in Figs. 1 (a) and 1 (c), but exactly corresponds to the case of Fig. 1 (b), further confirming the concept of decelerated generation and negligible $E'\leftarrow P_b$ conversion at \emph{low dose rate}.
The sublinear dependence of [$E’$] at relatively high dose rate is also totally different from the curves in Figs. 1 (a) and 1 (b), but exactly corresponds to the case of Fig. 1 (c).
This proves the presence of a remarkable backward conversion 
at \emph{high dose rate},
which provides a dominating FPL term in the dose profile.
In Fig. 2 (b) where [$P_b$] is plotted, it is seen that in both low or high dose rate, a super-linear and sublinear dose dependence is observed at small and large dose, respectively.
This is consistent with the cases of Figs. 1(e) and 1(f) but differs from the case of Fig. 1(d).
For $D>3D_c$, [$P_b$] tends to reach an asymptotic value of $P_b (D) = \lambda g_e D^\beta$.
This naturally explains a universal 2/3 power law dose dependence that has been experimentally observed for [$P_b$]~\cite{winokur1980interface}.

The individual values of [$E'$] and [$P_b$] can be fitted by our model with the already extracted $g_e$ and $\beta$.
A good match is found between the data and models in Fig. 2.
The experimental results also display clear sample-to-sample variability, but also can be perfectly fitted by the model, see Fig. S2 in the SM.
The extracted conversion rate constants
$\kappa_f$ and $\kappa_b$ are plotted in Fig. 3(c) as a function of dose rate.
It is seen that $\kappa_b$ (blue) is negligible at low dose rate, which is consistent with the large reaction barrier of $H^+ \leftarrow P_b$.
Both $\kappa_b$ and $\kappa_f$ increase as the dose rate increases.
This is a result of the increased carrier recombination.
From Fig. 3 (c), it is also noticed $\kappa_f$ increases slower than $\kappa_b$ as the dose rate increases.
This is consistent with the different barriers of the forward and backward reactions as analyzed before.
Accordingly, $\lambda$ show a negative $q$ dependence, as seen in Fig. S3 (a) in the SM.
The larger $\kappa_f$  
increases sublinearly with dose rate, hence $D_c=q/(\kappa_f+\kappa_b)$ increases slightly with dose rate, see Fig. S3 (b) in the SM.

\begin{figure}[!t]
  \includegraphics[width=\linewidth]{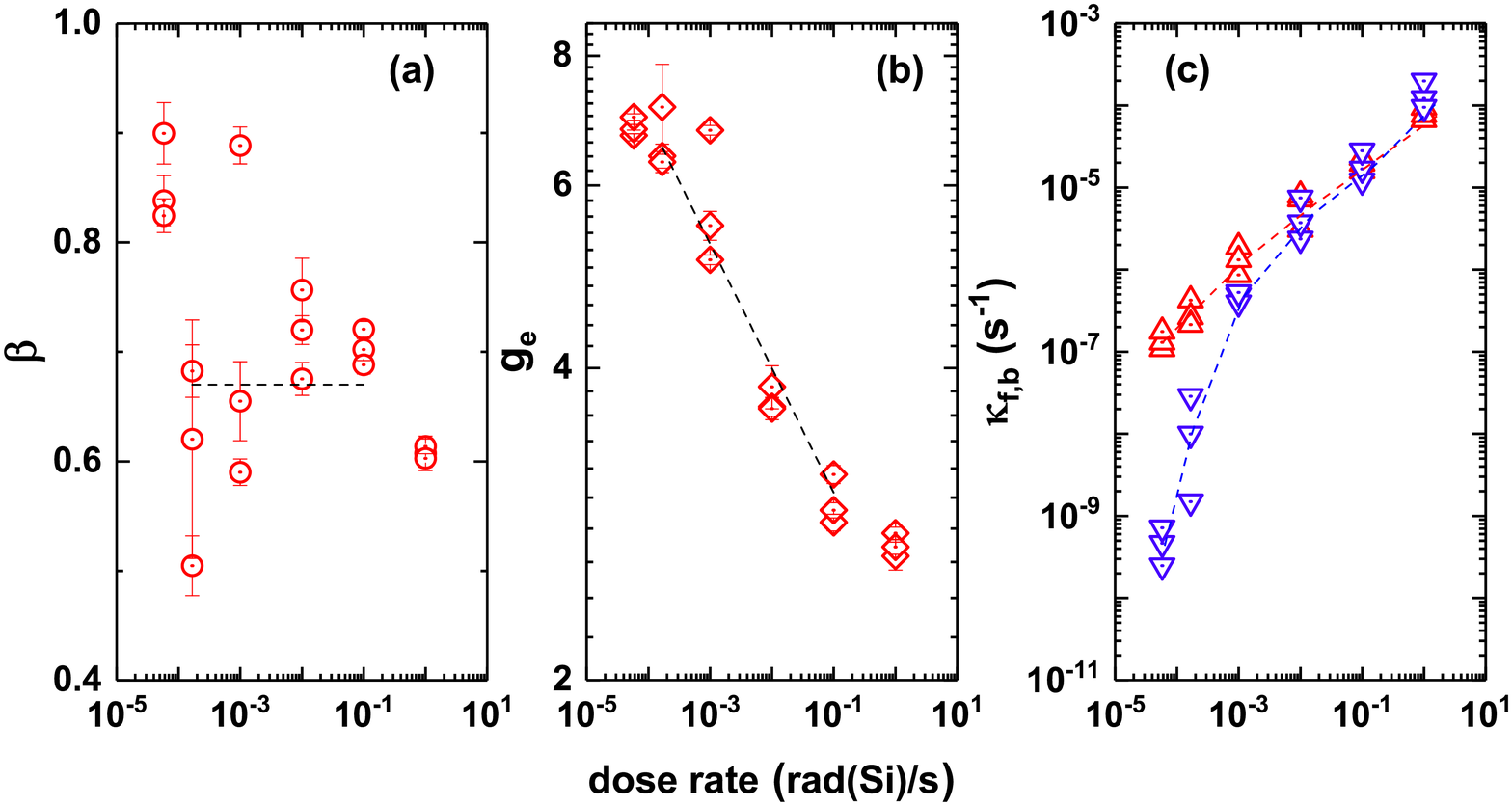}\\ 
  \caption{(a) $\beta$, (b) $g_e$, (c) $\kappa_f$ (red), and $\kappa_b$ (blue) from the fitting of the data as a function of dose rate.
}\label{ }
\end{figure}

For $D>3D_c$, [$P_b$]=$\lambda g_e D^\beta$ and the 
negative $q$ dependence of it at a fixed $D$ originates from the negative $q$ dependence of $g_e$ and $\lambda$, whose origins have been discussed above.
For $D<3D_c$, the dose rate dependence of $D_c$  
is also responsible for the ELDRS effect.
It would be interesting to compare our results with the mechanisms deduced from the traditional models.
The ``competition'' mechanism~\cite{boch2006dose}
suggests that a competition between hole trapping and recombination 
is the dominating reason for the $q$ dependence of the generation process, which would have a $q^{0.5-1}$ dependence due to an exponential distribution of the defect energy levels (giving $\sqrt{q}$)~\cite{fowler1956x,fowler1954conductivity-1} and the difference in the time cost at different dose rates (giving $q^{-1}$).
However, our results indicate that the dispersive transport of holes in disordered silica leads to a much gentle dependence of $q^{0.5-\beta}$, where the difference in the time costs 
is reduced to $q^{-(1-\alpha)}$. 
The ``hydrogen dimerization'' mechanism~\cite{hjalmarson2003mechanisms,rowsey2012mechanisms} suggests that a negative $q$ dependence of the conversion ratio can arise from the $E'\leftarrow H^+$ backward conversion at high dose rate, in which case, the protons pile up at the silica-silicon interface before reacting.
From Eq. (3) and the conversion rate equations, we obtain
[$\rm H^+$]=$\frac{k_{f1}{\rm [H_2]}-k_{b2}{\rm [H_2]}}
{(1-\lambda)^{-1} k_{b1}{\rm [EH]}- \lambda^{-1}k_{f2}{\rm [P_bH]}} g_e D^\beta$ for $D>3D_c$,
which indicates that the $H^+\leftarrow P_b$ backward conversion also contributes to the pile up of protons.
The ``space charge'' mechanism~\cite{fleetwood1994physical,rashkeev2002physical}
suggests that the negative $q$ dependence of the conversion ratio 
can also arise from the Coulomb repulsion of positive $E'$ on the transport of positive $\rm H^+$ to the interface.
However, 
our extracted $\kappa_f$ indicates that the transport and reaction of protons should be accelerated at high dose rate.

The proposed and verified analytic model, Eq. (3), indicates exactly what factors
in the silica-silicon structure influence the concentration of generated defects, so it can be used to guide the design of radiation-hard devices.
For example, [$E'+P_b$]$=k’P_0q^{0.5-\beta}D^\beta$.
So, it is crucial to decrease the concentration of $V_{O\gamma}$ ($P_0$) 
to reduce the total concentration of the irradiation-induced defects.
For fixed $k_{f,b}$'s, 
the conversion ratio $\lambda$ can be tuned by the ratio of [$\rm EH$]/[${\rm P_bH}$].
This means that, the weight of [$P_b$] ([$E'$]) in the total concentration can be reduced by increasing (decreasing) this ratio.

In summary, we have proposed a new irradiation-induced defect dynamics model in silica/silicon structures based on
two new concepts. 
The first concept is a decelerated generation of $E'$ centers in bulk silica, which originates from the dispersive diffusion of induced holes in disordered silica and is responsible for the general FPL dose dependence of [$E'$+$P_b$], the non-monotonic dose dependence of [$E'$] for low-dose-rate irradiation,
as well as the gentle dose rate dependence of the generation process.
The second concept is a reversible conversion between $E'$ and $P_b$ at the silica/silicon interface, which originates from the recombination enhanced reaction under irradiation and is indispensable for the sublinear dose dependence of [$E'$] for high-dose-rate irradiation as well as the negative dose rate dependence of the conversion process.
The derived analytical model based on these two new concepts can quantitatively describe the experimental data measured with wide dose and dose rate spectrum.
These remarkable results provide a solid foundation in the goal of understanding, predicting, and control the irradiation damage of silicon-based semiconductor devices.

This work was financially supported by the Science Challenge Project under Grant  Nos. TZ2016003 and TZ2018004, and NSFC under Grant Nos. 11991060, 12088101, 11634003,  and U1930402.


\begin{thebibliography}{60}%
\makeatletter
\providecommand \@ifxundefined [1]{%
 \@ifx{#1\undefined}
}%
\providecommand \@ifnum [1]{%
 \ifnum #1\expandafter \@firstoftwo
 \else \expandafter \@secondoftwo
 \fi
}%
\providecommand \@ifx [1]{%
 \ifx #1\expandafter \@firstoftwo
 \else \expandafter \@secondoftwo
 \fi
}%
\providecommand \natexlab [1]{#1}%
\providecommand \enquote  [1]{``#1''}%
\providecommand \bibnamefont  [1]{#1}%
\providecommand \bibfnamefont [1]{#1}%
\providecommand \citenamefont [1]{#1}%
\providecommand \href@noop [0]{\@secondoftwo}%
\providecommand \href [0]{\begingroup \@sanitize@url \@href}%
\providecommand \@href[1]{\@@startlink{#1}\@@href}%
\providecommand \@@href[1]{\endgroup#1\@@endlink}%
\providecommand \@sanitize@url [0]{\catcode `\\12\catcode `\$12\catcode
  `\&12\catcode `\#12\catcode `\^12\catcode `\_12\catcode `\%12\relax}%
\providecommand \@@startlink[1]{}%
\providecommand \@@endlink[0]{}%
\providecommand \url  [0]{\begingroup\@sanitize@url \@url }%
\providecommand \@url [1]{\endgroup\@href {#1}{\urlprefix }}%
\providecommand \urlprefix  [0]{URL }%
\providecommand \Eprint [0]{\href }%
\providecommand \doibase [0]{http://dx.doi.org/}%
\providecommand \selectlanguage [0]{\@gobble}%
\providecommand \bibinfo  [0]{\@secondoftwo}%
\providecommand \bibfield  [0]{\@secondoftwo}%
\providecommand \translation [1]{[#1]}%
\providecommand \BibitemOpen [0]{}%
\providecommand \bibitemStop [0]{}%
\providecommand \bibitemNoStop [0]{.\EOS\space}%
\providecommand \EOS [0]{\spacefactor3000\relax}%
\providecommand \BibitemShut  [1]{\csname bibitem#1\endcsname}%
\let\auto@bib@innerbib\@empty
\bibitem [{\citenamefont {Adell}\ and\ \citenamefont
  {Boch}(2014)}]{adell2014dose}%
  \BibitemOpen
  \bibfield  {author} {\bibinfo {author} {\bibfnamefont {P.}~\bibnamefont
  {Adell}}\ and\ \bibinfo {author} {\bibfnamefont {J.}~\bibnamefont {Boch}},\
  }\href@noop {} {\bibfield  {journal} {\bibinfo  {journal} {Proc NSREC Short
  Course}\ } (\bibinfo {year} {2014})}\BibitemShut {NoStop}%
\bibitem [{\citenamefont {Lenahan}\ and\ \citenamefont
  {Dressendorfer}(1984)}]{lenahan1984hole}%
  \BibitemOpen
  \bibfield  {author} {\bibinfo {author} {\bibfnamefont {P.~M.}\ \bibnamefont
  {Lenahan}}\ and\ \bibinfo {author} {\bibfnamefont {P.}~\bibnamefont
  {Dressendorfer}},\ }\href@noop {} {\bibfield  {journal} {\bibinfo  {journal}
  {J. Appl. Phys.}\ }\textbf {\bibinfo {volume} {55}},\ \bibinfo {pages} {3495}
  (\bibinfo {year} {1984})}\BibitemShut {NoStop}%
\bibitem [{\citenamefont {Poindexter}\ \emph {et~al.}(1984)\citenamefont
  {Poindexter}, \citenamefont {Gerardi}, \citenamefont {Rueckel}, \citenamefont
  {Caplan}, \citenamefont {Johnson},\ and\ \citenamefont
  {Biegelsen}}]{poindexter1984electronic}%
  \BibitemOpen
  \bibfield  {author} {\bibinfo {author} {\bibfnamefont {E.}~\bibnamefont
  {Poindexter}}, \bibinfo {author} {\bibfnamefont {G.}~\bibnamefont {Gerardi}},
  \bibinfo {author} {\bibfnamefont {M.-E.}\ \bibnamefont {Rueckel}}, \bibinfo
  {author} {\bibfnamefont {P.}~\bibnamefont {Caplan}}, \bibinfo {author}
  {\bibfnamefont {N.}~\bibnamefont {Johnson}}, \ and\ \bibinfo {author}
  {\bibfnamefont {D.}~\bibnamefont {Biegelsen}},\ }\href@noop {} {\bibfield
  {journal} {\bibinfo  {journal} {J. Appl. Phys.}\ }\textbf {\bibinfo {volume}
  {56}},\ \bibinfo {pages} {2844} (\bibinfo {year} {1984})}\BibitemShut
  {NoStop}%
\bibitem [{\citenamefont {Weeks}(1956)}]{weeks1956paramagnetic}%
  \BibitemOpen
  \bibfield  {author} {\bibinfo {author} {\bibfnamefont {R.}~\bibnamefont
  {Weeks}},\ }\href@noop {} {\bibfield  {journal} {\bibinfo  {journal} {J.
  Appl. Phys.}\ }\textbf {\bibinfo {volume} {27}},\ \bibinfo {pages} {1376}
  (\bibinfo {year} {1956})}\BibitemShut {NoStop}%
\bibitem [{\citenamefont {Poindexter}\ \emph {et~al.}(1981)\citenamefont
  {Poindexter}, \citenamefont {Caplan}, \citenamefont {Deal},\ and\
  \citenamefont {Razouk}}]{poindexter1981interface}%
  \BibitemOpen
  \bibfield  {author} {\bibinfo {author} {\bibfnamefont {E.~H.}\ \bibnamefont
  {Poindexter}}, \bibinfo {author} {\bibfnamefont {P.~J.}\ \bibnamefont
  {Caplan}}, \bibinfo {author} {\bibfnamefont {B.~E.}\ \bibnamefont {Deal}}, \
  and\ \bibinfo {author} {\bibfnamefont {R.~R.}\ \bibnamefont {Razouk}},\
  }\href@noop {} {\bibfield  {journal} {\bibinfo  {journal} {J. Appl. Phys.}\
  }\textbf {\bibinfo {volume} {52}},\ \bibinfo {pages} {879} (\bibinfo {year}
  {1981})}\BibitemShut {NoStop}%
\bibitem [{\citenamefont {Fleetwood}\ \emph {et~al.}(1993)\citenamefont
  {Fleetwood}, \citenamefont {Winokur}, \citenamefont {Reber~Jr}, \citenamefont
  {Meisenheimer}, \citenamefont {Schwank}, \citenamefont {Shaneyfelt},\ and\
  \citenamefont {Riewe}}]{fleetwood1993effects}%
  \BibitemOpen
  \bibfield  {author} {\bibinfo {author} {\bibfnamefont {D.}~\bibnamefont
  {Fleetwood}}, \bibinfo {author} {\bibfnamefont {P.}~\bibnamefont {Winokur}},
  \bibinfo {author} {\bibfnamefont {R.}~\bibnamefont {Reber~Jr}}, \bibinfo
  {author} {\bibfnamefont {T.}~\bibnamefont {Meisenheimer}}, \bibinfo {author}
  {\bibfnamefont {J.}~\bibnamefont {Schwank}}, \bibinfo {author} {\bibfnamefont
  {M.}~\bibnamefont {Shaneyfelt}}, \ and\ \bibinfo {author} {\bibfnamefont
  {L.}~\bibnamefont {Riewe}},\ }\href@noop {} {\bibfield  {journal} {\bibinfo
  {journal} {J. Appl. Phys.}\ }\textbf {\bibinfo {volume} {73}},\ \bibinfo
  {pages} {5058} (\bibinfo {year} {1993})}\BibitemShut {NoStop}%
\bibitem [{\citenamefont {McLean}(1980)}]{mclean1980framework}%
  \BibitemOpen
  \bibfield  {author} {\bibinfo {author} {\bibfnamefont {F.}~\bibnamefont
  {McLean}},\ }\href@noop {} {\bibfield  {journal} {\bibinfo  {journal} {IEEE
  Trans. Nucl. Sci.}\ }\textbf {\bibinfo {volume} {27}},\ \bibinfo {pages}
  {1651} (\bibinfo {year} {1980})}\BibitemShut {NoStop}%
\bibitem [{\citenamefont {Oldham}\ and\ \citenamefont
  {McLean}(2003)}]{oldham2003total}%
  \BibitemOpen
  \bibfield  {author} {\bibinfo {author} {\bibfnamefont {T.~R.}\ \bibnamefont
  {Oldham}}\ and\ \bibinfo {author} {\bibfnamefont {F.}~\bibnamefont
  {McLean}},\ }\href@noop {} {\bibfield  {journal} {\bibinfo  {journal} {IEEE
  Trans. Nucl. Sci.}\ }\textbf {\bibinfo {volume} {50}},\ \bibinfo {pages}
  {483} (\bibinfo {year} {2003})}\BibitemShut {NoStop}%
\bibitem [{\citenamefont {Fleetwood}(2017)}]{fleetwood2017evolution}%
  \BibitemOpen
  \bibfield  {author} {\bibinfo {author} {\bibfnamefont {D.~M.}\ \bibnamefont
  {Fleetwood}},\ }\href@noop {} {\bibfield  {journal} {\bibinfo  {journal}
  {IEEE Trans. Nucl. Sci.}\ }\textbf {\bibinfo {volume} {65}},\ \bibinfo
  {pages} {1465} (\bibinfo {year} {2017})}\BibitemShut {NoStop}%
\bibitem [{\citenamefont {Shaneyfelt}\ \emph {et~al.}(1991)\citenamefont
  {Shaneyfelt}, \citenamefont {Fleetwood}, \citenamefont {Schwank},\ and\
  \citenamefont {Hughes}}]{Shaneyfelt1991charge}%
  \BibitemOpen
  \bibfield  {author} {\bibinfo {author} {\bibfnamefont {M.}~\bibnamefont
  {Shaneyfelt}}, \bibinfo {author} {\bibfnamefont {D.}~\bibnamefont
  {Fleetwood}}, \bibinfo {author} {\bibfnamefont {J.}~\bibnamefont {Schwank}},
  \ and\ \bibinfo {author} {\bibfnamefont {K.}~\bibnamefont {Hughes}},\
  }\href@noop {} {\bibfield  {journal} {\bibinfo  {journal} {IEEE Trans. Nucl.
  Sci.}\ }\textbf {\bibinfo {volume} {38}},\ \bibinfo {pages} {1187} (\bibinfo
  {year} {1991})}\BibitemShut {NoStop}%
\bibitem [{\citenamefont {Boch}\ \emph
  {et~al.}(2006{\natexlab{a}})\citenamefont {Boch}, \citenamefont {Saigne},
  \citenamefont {Dusseau},\ and\ \citenamefont
  {Schrimpf}}]{boch2006temperature}%
  \BibitemOpen
  \bibfield  {author} {\bibinfo {author} {\bibfnamefont {J.}~\bibnamefont
  {Boch}}, \bibinfo {author} {\bibfnamefont {F.}~\bibnamefont {Saigne}},
  \bibinfo {author} {\bibfnamefont {L.}~\bibnamefont {Dusseau}}, \ and\
  \bibinfo {author} {\bibfnamefont {R.}~\bibnamefont {Schrimpf}},\ }\href@noop
  {} {\bibfield  {journal} {\bibinfo  {journal} {Appl. Phys. Lett.}\ }\textbf
  {\bibinfo {volume} {89}},\ \bibinfo {pages} {042108} (\bibinfo {year}
  {2006}{\natexlab{a}})}\BibitemShut {NoStop}%
\bibitem [{\citenamefont {Hughes}(1977)}]{hughes1977time}%
  \BibitemOpen
  \bibfield  {author} {\bibinfo {author} {\bibfnamefont {R.}~\bibnamefont
  {Hughes}},\ }\href@noop {} {\bibfield  {journal} {\bibinfo  {journal} {Phys.
  Rev. B}\ }\textbf {\bibinfo {volume} {15}},\ \bibinfo {pages} {2012}
  (\bibinfo {year} {1977})}\BibitemShut {NoStop}%
\bibitem [{\citenamefont {Boesch}\ \emph {et~al.}(1975)\citenamefont {Boesch},
  \citenamefont {McLean}, \citenamefont {McGarrity},\ and\ \citenamefont
  {Ausman}}]{boesch1975hole}%
  \BibitemOpen
  \bibfield  {author} {\bibinfo {author} {\bibfnamefont {H.}~\bibnamefont
  {Boesch}}, \bibinfo {author} {\bibfnamefont {F.}~\bibnamefont {McLean}},
  \bibinfo {author} {\bibfnamefont {J.}~\bibnamefont {McGarrity}}, \ and\
  \bibinfo {author} {\bibfnamefont {G.}~\bibnamefont {Ausman}},\ }\href@noop {}
  {\bibfield  {journal} {\bibinfo  {journal} {IEEE Trans. Nucl. Sci.}\ }\textbf
  {\bibinfo {volume} {22}},\ \bibinfo {pages} {2163} (\bibinfo {year}
  {1975})}\BibitemShut {NoStop}%
\bibitem [{\citenamefont {Nicklaw}\ \emph {et~al.}(2002)\citenamefont
  {Nicklaw}, \citenamefont {Lu}, \citenamefont {Fleetwood}, \citenamefont
  {Schrimpf},\ and\ \citenamefont {Pantelides}}]{nicklaw2002structure}%
  \BibitemOpen
  \bibfield  {author} {\bibinfo {author} {\bibfnamefont {C.}~\bibnamefont
  {Nicklaw}}, \bibinfo {author} {\bibfnamefont {Z.-Y.}\ \bibnamefont {Lu}},
  \bibinfo {author} {\bibfnamefont {D.}~\bibnamefont {Fleetwood}}, \bibinfo
  {author} {\bibfnamefont {R.}~\bibnamefont {Schrimpf}}, \ and\ \bibinfo
  {author} {\bibfnamefont {S.}~\bibnamefont {Pantelides}},\ }\href@noop {}
  {\bibfield  {journal} {\bibinfo  {journal} {IEEE Trans. Nucl. Sci.}\ }\textbf
  {\bibinfo {volume} {49}},\ \bibinfo {pages} {2667} (\bibinfo {year}
  {2002})}\BibitemShut {NoStop}%
\bibitem [{\citenamefont {Conley}\ and\ \citenamefont
  {Lenahan}(1993{\natexlab{a}})}]{conley1993room}%
  \BibitemOpen
  \bibfield  {author} {\bibinfo {author} {\bibfnamefont {J.}~\bibnamefont
  {Conley}}\ and\ \bibinfo {author} {\bibfnamefont {P.}~\bibnamefont
  {Lenahan}},\ }\href@noop {} {\bibfield  {journal} {\bibinfo  {journal} {Appl.
  Phys. Lett.}\ }\textbf {\bibinfo {volume} {62}},\ \bibinfo {pages} {40}
  (\bibinfo {year} {1993}{\natexlab{a}})}\BibitemShut {NoStop}%
\bibitem [{\citenamefont {Conley~Jr}\ \emph
  {et~al.}(1994{\natexlab{a}})\citenamefont {Conley~Jr}, \citenamefont
  {Lenahan}, \citenamefont {Evans}, \citenamefont {Lowry},\ and\ \citenamefont
  {Morthorst}}]{conley1994observation_JAP}%
  \BibitemOpen
  \bibfield  {author} {\bibinfo {author} {\bibfnamefont {J.~F.}\ \bibnamefont
  {Conley~Jr}}, \bibinfo {author} {\bibfnamefont {P.~M.}\ \bibnamefont
  {Lenahan}}, \bibinfo {author} {\bibfnamefont {H.}~\bibnamefont {Evans}},
  \bibinfo {author} {\bibfnamefont {R.}~\bibnamefont {Lowry}}, \ and\ \bibinfo
  {author} {\bibfnamefont {T.}~\bibnamefont {Morthorst}},\ }\href@noop {}
  {\bibfield  {journal} {\bibinfo  {journal} {J. Appl. Phys.}\ }\textbf
  {\bibinfo {volume} {76}},\ \bibinfo {pages} {2872} (\bibinfo {year}
  {1994}{\natexlab{a}})}\BibitemShut {NoStop}%
\bibitem [{\citenamefont {Bl{\"o}chl}(2000)}]{blochl2000first}%
  \BibitemOpen
  \bibfield  {author} {\bibinfo {author} {\bibfnamefont {P.~E.}\ \bibnamefont
  {Bl{\"o}chl}},\ }\href@noop {} {\bibfield  {journal} {\bibinfo  {journal}
  {Phys. Rev. B}\ }\textbf {\bibinfo {volume} {62}},\ \bibinfo {pages} {6158}
  (\bibinfo {year} {2000})}\BibitemShut {NoStop}%
\bibitem [{\citenamefont {Lu}\ \emph {et~al.}(2002)\citenamefont {Lu},
  \citenamefont {Nicklaw}, \citenamefont {Fleetwood}, \citenamefont
  {Schrimpf},\ and\ \citenamefont {Pantelides}}]{lu2002structure}%
  \BibitemOpen
  \bibfield  {author} {\bibinfo {author} {\bibfnamefont {Z.-Y.}\ \bibnamefont
  {Lu}}, \bibinfo {author} {\bibfnamefont {C.}~\bibnamefont {Nicklaw}},
  \bibinfo {author} {\bibfnamefont {D.}~\bibnamefont {Fleetwood}}, \bibinfo
  {author} {\bibfnamefont {R.}~\bibnamefont {Schrimpf}}, \ and\ \bibinfo
  {author} {\bibfnamefont {S.}~\bibnamefont {Pantelides}},\ }\href@noop {}
  {\bibfield  {journal} {\bibinfo  {journal} {Phys. Rev. Lett.}\ }\textbf
  {\bibinfo {volume} {89}},\ \bibinfo {pages} {285505} (\bibinfo {year}
  {2002})}\BibitemShut {NoStop}%
\bibitem [{\citenamefont {Conley~Jr}\ \emph
  {et~al.}(1994{\natexlab{b}})\citenamefont {Conley~Jr}, \citenamefont
  {Lenahan}, \citenamefont {Evans}, \citenamefont {Lowry},\ and\ \citenamefont
  {Morthorst}}]{conley1994observation}%
  \BibitemOpen
  \bibfield  {author} {\bibinfo {author} {\bibfnamefont {J.~F.}\ \bibnamefont
  {Conley~Jr}}, \bibinfo {author} {\bibfnamefont {P.}~\bibnamefont {Lenahan}},
  \bibinfo {author} {\bibfnamefont {H.}~\bibnamefont {Evans}}, \bibinfo
  {author} {\bibfnamefont {R.}~\bibnamefont {Lowry}}, \ and\ \bibinfo {author}
  {\bibfnamefont {T.}~\bibnamefont {Morthorst}},\ }\href@noop {} {\bibfield
  {journal} {\bibinfo  {journal} {Appl. Phys. Lett.}\ }\textbf {\bibinfo
  {volume} {65}},\ \bibinfo {pages} {2281} (\bibinfo {year}
  {1994}{\natexlab{b}})}\BibitemShut {NoStop}%
\bibitem [{\citenamefont {Yang}\ \emph {et~al.}(2016)\citenamefont {Yang},
  \citenamefont {Shi}, \citenamefont {Wang},\ and\ \citenamefont
  {Wei}}]{yang2016non}%
  \BibitemOpen
  \bibfield  {author} {\bibinfo {author} {\bibfnamefont {J.-H.}\ \bibnamefont
  {Yang}}, \bibinfo {author} {\bibfnamefont {L.}~\bibnamefont {Shi}}, \bibinfo
  {author} {\bibfnamefont {L.-W.}\ \bibnamefont {Wang}}, \ and\ \bibinfo
  {author} {\bibfnamefont {S.-H.}\ \bibnamefont {Wei}},\ }\href@noop {}
  {\bibfield  {journal} {\bibinfo  {journal} {Sci. Rep.}\ }\textbf {\bibinfo
  {volume} {6}},\ \bibinfo {pages} {1} (\bibinfo {year} {2016})}\BibitemShut
  {NoStop}%
\bibitem [{\citenamefont {Conley}\ and\ \citenamefont
  {Lenahan}(1993{\natexlab{b}})}]{conley1993molecular}%
  \BibitemOpen
  \bibfield  {author} {\bibinfo {author} {\bibfnamefont {J.}~\bibnamefont
  {Conley}}\ and\ \bibinfo {author} {\bibfnamefont {P.}~\bibnamefont
  {Lenahan}},\ }\href@noop {} {\bibfield  {journal} {\bibinfo  {journal} {IEEE
  Trans. Nucl. Sci.}\ }\textbf {\bibinfo {volume} {40}},\ \bibinfo {pages}
  {1335} (\bibinfo {year} {1993}{\natexlab{b}})}\BibitemShut {NoStop}%
\bibitem [{\citenamefont {Stahlbush}\ \emph {et~al.}(1993)\citenamefont
  {Stahlbush}, \citenamefont {Edwards}, \citenamefont {Griscom},\ and\
  \citenamefont {Mrstik}}]{stahlbush1993post}%
  \BibitemOpen
  \bibfield  {author} {\bibinfo {author} {\bibfnamefont {R.}~\bibnamefont
  {Stahlbush}}, \bibinfo {author} {\bibfnamefont {A.}~\bibnamefont {Edwards}},
  \bibinfo {author} {\bibfnamefont {D.}~\bibnamefont {Griscom}}, \ and\
  \bibinfo {author} {\bibfnamefont {B.}~\bibnamefont {Mrstik}},\ }\href@noop {}
  {\bibfield  {journal} {\bibinfo  {journal} {J. Appl. Phys.}\ }\textbf
  {\bibinfo {volume} {73}},\ \bibinfo {pages} {658} (\bibinfo {year}
  {1993})}\BibitemShut {NoStop}%
\bibitem [{\citenamefont {Tuttle}\ \emph {et~al.}(2010)\citenamefont {Tuttle},
  \citenamefont {Hughart}, \citenamefont {Schrimpf}, \citenamefont
  {Fleetwood},\ and\ \citenamefont {Pantelides}}]{tuttle2010defect}%
  \BibitemOpen
  \bibfield  {author} {\bibinfo {author} {\bibfnamefont {B.~R.}\ \bibnamefont
  {Tuttle}}, \bibinfo {author} {\bibfnamefont {D.~R.}\ \bibnamefont {Hughart}},
  \bibinfo {author} {\bibfnamefont {R.~D.}\ \bibnamefont {Schrimpf}}, \bibinfo
  {author} {\bibfnamefont {D.~M.}\ \bibnamefont {Fleetwood}}, \ and\ \bibinfo
  {author} {\bibfnamefont {S.~T.}\ \bibnamefont {Pantelides}},\ }\href@noop {}
  {\bibfield  {journal} {\bibinfo  {journal} {IEEE Trans. Nucl. Sci.}\ }\textbf
  {\bibinfo {volume} {57}},\ \bibinfo {pages} {3046} (\bibinfo {year}
  {2010})}\BibitemShut {NoStop}%
\bibitem [{\citenamefont {Mrstik}\ and\ \citenamefont
  {Rendell}(1991)}]{mrstik1991si}%
  \BibitemOpen
  \bibfield  {author} {\bibinfo {author} {\bibfnamefont {B.}~\bibnamefont
  {Mrstik}}\ and\ \bibinfo {author} {\bibfnamefont {R.}~\bibnamefont
  {Rendell}},\ }\href@noop {} {\bibfield  {journal} {\bibinfo  {journal} {IEEE
  Trans. Nucl. Sci.}\ }\textbf {\bibinfo {volume} {38}},\ \bibinfo {pages}
  {1101} (\bibinfo {year} {1991})}\BibitemShut {NoStop}%
\bibitem [{\citenamefont {Stathis}\ and\ \citenamefont
  {Cartier}(1994)}]{stathis1994atomic}%
  \BibitemOpen
  \bibfield  {author} {\bibinfo {author} {\bibfnamefont {J.}~\bibnamefont
  {Stathis}}\ and\ \bibinfo {author} {\bibfnamefont {E.}~\bibnamefont
  {Cartier}},\ }\href@noop {} {\bibfield  {journal} {\bibinfo  {journal} {Phys.
  Rev. Lett.}\ }\textbf {\bibinfo {volume} {72}},\ \bibinfo {pages} {2745}
  (\bibinfo {year} {1994})}\BibitemShut {NoStop}%
\bibitem [{\citenamefont {Rashkeev}\ \emph {et~al.}(2001)\citenamefont
  {Rashkeev}, \citenamefont {Fleetwood}, \citenamefont {Schrimpf},\ and\
  \citenamefont {Pantelides}}]{rashkeev2001defect}%
  \BibitemOpen
  \bibfield  {author} {\bibinfo {author} {\bibfnamefont {S.}~\bibnamefont
  {Rashkeev}}, \bibinfo {author} {\bibfnamefont {D.}~\bibnamefont {Fleetwood}},
  \bibinfo {author} {\bibfnamefont {R.}~\bibnamefont {Schrimpf}}, \ and\
  \bibinfo {author} {\bibfnamefont {S.}~\bibnamefont {Pantelides}},\
  }\href@noop {} {\bibfield  {journal} {\bibinfo  {journal} {Phys. Rev. Lett.}\
  }\textbf {\bibinfo {volume} {87}},\ \bibinfo {pages} {165506} (\bibinfo
  {year} {2001})}\BibitemShut {NoStop}%
\bibitem [{\citenamefont {Hjalmarson}\ \emph {et~al.}(2003)\citenamefont
  {Hjalmarson}, \citenamefont {Pease}, \citenamefont {Witczak}, \citenamefont
  {Shaneyfelt}, \citenamefont {Schwank}, \citenamefont {Edwards}, \citenamefont
  {Hembree},\ and\ \citenamefont {Mattsson}}]{hjalmarson2003mechanisms}%
  \BibitemOpen
  \bibfield  {author} {\bibinfo {author} {\bibfnamefont {H.~P.}\ \bibnamefont
  {Hjalmarson}}, \bibinfo {author} {\bibfnamefont {R.~L.}\ \bibnamefont
  {Pease}}, \bibinfo {author} {\bibfnamefont {S.~C.}\ \bibnamefont {Witczak}},
  \bibinfo {author} {\bibfnamefont {M.~R.}\ \bibnamefont {Shaneyfelt}},
  \bibinfo {author} {\bibfnamefont {J.~R.}\ \bibnamefont {Schwank}}, \bibinfo
  {author} {\bibfnamefont {A.~H.}\ \bibnamefont {Edwards}}, \bibinfo {author}
  {\bibfnamefont {C.~E.}\ \bibnamefont {Hembree}}, \ and\ \bibinfo {author}
  {\bibfnamefont {T.~R.}\ \bibnamefont {Mattsson}},\ }\href@noop {} {\bibfield
  {journal} {\bibinfo  {journal} {IEEE Trans. Nucl. Sci.}\ }\textbf {\bibinfo
  {volume} {50}},\ \bibinfo {pages} {1901} (\bibinfo {year}
  {2003})}\BibitemShut {NoStop}%
\bibitem [{\citenamefont {Rowsey}\ \emph {et~al.}(2011)\citenamefont {Rowsey},
  \citenamefont {Law}, \citenamefont {Schrimpf}, \citenamefont {Fleetwood},
  \citenamefont {Tuttle},\ and\ \citenamefont
  {Pantelides}}]{rowsey2011quantitative}%
  \BibitemOpen
  \bibfield  {author} {\bibinfo {author} {\bibfnamefont {N.~L.}\ \bibnamefont
  {Rowsey}}, \bibinfo {author} {\bibfnamefont {M.~E.}\ \bibnamefont {Law}},
  \bibinfo {author} {\bibfnamefont {R.~D.}\ \bibnamefont {Schrimpf}}, \bibinfo
  {author} {\bibfnamefont {D.~M.}\ \bibnamefont {Fleetwood}}, \bibinfo {author}
  {\bibfnamefont {B.~R.}\ \bibnamefont {Tuttle}}, \ and\ \bibinfo {author}
  {\bibfnamefont {S.~T.}\ \bibnamefont {Pantelides}},\ }\href@noop {}
  {\bibfield  {journal} {\bibinfo  {journal} {IEEE Trans. Nucl. Sci.}\ }\textbf
  {\bibinfo {volume} {58}},\ \bibinfo {pages} {2937} (\bibinfo {year}
  {2011})}\BibitemShut {NoStop}%
\bibitem [{\citenamefont {Rowsey}\ \emph
  {et~al.}(2012{\natexlab{a}})\citenamefont {Rowsey}, \citenamefont {Law},
  \citenamefont {Schrimpf}, \citenamefont {Fleetwood}, \citenamefont {Tuttle},\
  and\ \citenamefont {Pantelides}}]{rowsey2012radiation}%
  \BibitemOpen
  \bibfield  {author} {\bibinfo {author} {\bibfnamefont {N.~L.}\ \bibnamefont
  {Rowsey}}, \bibinfo {author} {\bibfnamefont {M.~E.}\ \bibnamefont {Law}},
  \bibinfo {author} {\bibfnamefont {R.~D.}\ \bibnamefont {Schrimpf}}, \bibinfo
  {author} {\bibfnamefont {D.~M.}\ \bibnamefont {Fleetwood}}, \bibinfo {author}
  {\bibfnamefont {B.~R.}\ \bibnamefont {Tuttle}}, \ and\ \bibinfo {author}
  {\bibfnamefont {S.~T.}\ \bibnamefont {Pantelides}},\ }\href@noop {}
  {\bibfield  {journal} {\bibinfo  {journal} {IEEE Trans. Nucl. Sci.}\ }\textbf
  {\bibinfo {volume} {59}},\ \bibinfo {pages} {755} (\bibinfo {year}
  {2012}{\natexlab{a}})}\BibitemShut {NoStop}%
\bibitem [{\citenamefont {Rowsey}\ \emph
  {et~al.}(2012{\natexlab{b}})\citenamefont {Rowsey}, \citenamefont {Law},
  \citenamefont {Schrimpf}, \citenamefont {Fleetwood}, \citenamefont {Tuttle},\
  and\ \citenamefont {Pantelides}}]{rowsey2012mechanisms}%
  \BibitemOpen
  \bibfield  {author} {\bibinfo {author} {\bibfnamefont {N.~L.}\ \bibnamefont
  {Rowsey}}, \bibinfo {author} {\bibfnamefont {M.~E.}\ \bibnamefont {Law}},
  \bibinfo {author} {\bibfnamefont {R.~D.}\ \bibnamefont {Schrimpf}}, \bibinfo
  {author} {\bibfnamefont {D.~M.}\ \bibnamefont {Fleetwood}}, \bibinfo {author}
  {\bibfnamefont {B.~R.}\ \bibnamefont {Tuttle}}, \ and\ \bibinfo {author}
  {\bibfnamefont {S.~T.}\ \bibnamefont {Pantelides}},\ }\href@noop {}
  {\bibfield  {journal} {\bibinfo  {journal} {IEEE Trans. Nucl. Sci.}\ }\textbf
  {\bibinfo {volume} {59}},\ \bibinfo {pages} {3069} (\bibinfo {year}
  {2012}{\natexlab{b}})}\BibitemShut {NoStop}%
\bibitem [{\citenamefont {Hughart}\ \emph {et~al.}(2012)\citenamefont
  {Hughart}, \citenamefont {Schrimpf}, \citenamefont {Fleetwood}, \citenamefont
  {Rowsey}, \citenamefont {Law}, \citenamefont {Tuttle},\ and\ \citenamefont
  {Pantelides}}]{hughart2012effects}%
  \BibitemOpen
  \bibfield  {author} {\bibinfo {author} {\bibfnamefont {D.}~\bibnamefont
  {Hughart}}, \bibinfo {author} {\bibfnamefont {R.}~\bibnamefont {Schrimpf}},
  \bibinfo {author} {\bibfnamefont {D.}~\bibnamefont {Fleetwood}}, \bibinfo
  {author} {\bibfnamefont {N.}~\bibnamefont {Rowsey}}, \bibinfo {author}
  {\bibfnamefont {M.}~\bibnamefont {Law}}, \bibinfo {author} {\bibfnamefont
  {B.}~\bibnamefont {Tuttle}}, \ and\ \bibinfo {author} {\bibfnamefont
  {S.}~\bibnamefont {Pantelides}},\ }\href@noop {} {\bibfield  {journal}
  {\bibinfo  {journal} {IEEE Trans. Nucl. Sci.}\ }\textbf {\bibinfo {volume}
  {59}},\ \bibinfo {pages} {3087} (\bibinfo {year} {2012})}\BibitemShut
  {NoStop}%
\bibitem [{\citenamefont {Fahey}\ \emph {et~al.}(1989)\citenamefont {Fahey},
  \citenamefont {Griffin},\ and\ \citenamefont {Plummer}}]{fahey1989point}%
  \BibitemOpen
  \bibfield  {author} {\bibinfo {author} {\bibfnamefont {P.~M.}\ \bibnamefont
  {Fahey}}, \bibinfo {author} {\bibfnamefont {P.}~\bibnamefont {Griffin}}, \
  and\ \bibinfo {author} {\bibfnamefont {J.}~\bibnamefont {Plummer}},\
  }\href@noop {} {\bibfield  {journal} {\bibinfo  {journal} {Rev. Mod. Phys.}\
  }\textbf {\bibinfo {volume} {61}},\ \bibinfo {pages} {289} (\bibinfo {year}
  {1989})}\BibitemShut {NoStop}%
\bibitem [{\citenamefont {Li}\ \emph {et~al.}(2019)\citenamefont {Li},
  \citenamefont {Yang}, \citenamefont {Chen}, \citenamefont {Dong},\ and\
  \citenamefont {Lv}}]{li2019correlation}%
  \BibitemOpen
  \bibfield  {author} {\bibinfo {author} {\bibfnamefont {X.}~\bibnamefont
  {Li}}, \bibinfo {author} {\bibfnamefont {J.}~\bibnamefont {Yang}}, \bibinfo
  {author} {\bibfnamefont {H.}~\bibnamefont {Chen}}, \bibinfo {author}
  {\bibfnamefont {S.}~\bibnamefont {Dong}}, \ and\ \bibinfo {author}
  {\bibfnamefont {G.}~\bibnamefont {Lv}},\ }\href@noop {} {\bibfield  {journal}
  {\bibinfo  {journal} {IEEE Trans. Nucl. Sci.}\ }\textbf {\bibinfo {volume}
  {66}},\ \bibinfo {pages} {1612} (\bibinfo {year} {2019})}\BibitemShut
  {NoStop}%
\bibitem [{\citenamefont {Enlow}\ \emph {et~al.}(1991)\citenamefont {Enlow},
  \citenamefont {Pease}, \citenamefont {Combs},\ and\ \citenamefont
  {Schrimpf}}]{Enlow1991Response}%
  \BibitemOpen
  \bibfield  {author} {\bibinfo {author} {\bibfnamefont {E.~W.}\ \bibnamefont
  {Enlow}}, \bibinfo {author} {\bibfnamefont {R.~L.}\ \bibnamefont {Pease}},
  \bibinfo {author} {\bibfnamefont {W.}~\bibnamefont {Combs}}, \ and\ \bibinfo
  {author} {\bibfnamefont {R.~D.}\ \bibnamefont {Schrimpf}},\ }\href@noop {}
  {\bibfield  {journal} {\bibinfo  {journal} {IEEE Trans. Nucl. Sci.}\ }\textbf
  {\bibinfo {volume} {38}},\ \bibinfo {pages} {1342} (\bibinfo {year}
  {1991})}\BibitemShut {NoStop}%
\bibitem [{\citenamefont {Fleetwood}\ \emph {et~al.}(1994)\citenamefont
  {Fleetwood}, \citenamefont {Kosier}, \citenamefont {Nowlin}, \citenamefont
  {Schrimpf}, \citenamefont {Reber}, \citenamefont {DeLaus}, \citenamefont
  {Winokur}, \citenamefont {Wei}, \citenamefont {Combs},\ and\ \citenamefont
  {Pease}}]{fleetwood1994physical}%
  \BibitemOpen
  \bibfield  {author} {\bibinfo {author} {\bibfnamefont {D.}~\bibnamefont
  {Fleetwood}}, \bibinfo {author} {\bibfnamefont {S.}~\bibnamefont {Kosier}},
  \bibinfo {author} {\bibfnamefont {R.}~\bibnamefont {Nowlin}}, \bibinfo
  {author} {\bibfnamefont {R.}~\bibnamefont {Schrimpf}}, \bibinfo {author}
  {\bibfnamefont {R.}~\bibnamefont {Reber}}, \bibinfo {author} {\bibfnamefont
  {M.}~\bibnamefont {DeLaus}}, \bibinfo {author} {\bibfnamefont
  {P.}~\bibnamefont {Winokur}}, \bibinfo {author} {\bibfnamefont
  {A.}~\bibnamefont {Wei}}, \bibinfo {author} {\bibfnamefont {W.}~\bibnamefont
  {Combs}}, \ and\ \bibinfo {author} {\bibfnamefont {R.}~\bibnamefont
  {Pease}},\ }\href@noop {} {\bibfield  {journal} {\bibinfo  {journal} {IEEE
  Trans. Nucl. Sci.}\ }\textbf {\bibinfo {volume} {41}},\ \bibinfo {pages}
  {1871} (\bibinfo {year} {1994})}\BibitemShut {NoStop}%
\bibitem [{\citenamefont {Rashkeev}\ \emph {et~al.}(2002)\citenamefont
  {Rashkeev}, \citenamefont {Cirba}, \citenamefont {Fleetwood}, \citenamefont
  {Schrimpf}, \citenamefont {Witczak}, \citenamefont {Michez},\ and\
  \citenamefont {Pantelides}}]{rashkeev2002physical}%
  \BibitemOpen
  \bibfield  {author} {\bibinfo {author} {\bibfnamefont {S.~N.}\ \bibnamefont
  {Rashkeev}}, \bibinfo {author} {\bibfnamefont {C.~R.}\ \bibnamefont {Cirba}},
  \bibinfo {author} {\bibfnamefont {D.~M.}\ \bibnamefont {Fleetwood}}, \bibinfo
  {author} {\bibfnamefont {R.~D.}\ \bibnamefont {Schrimpf}}, \bibinfo {author}
  {\bibfnamefont {S.~C.}\ \bibnamefont {Witczak}}, \bibinfo {author}
  {\bibfnamefont {A.}~\bibnamefont {Michez}}, \ and\ \bibinfo {author}
  {\bibfnamefont {S.~T.}\ \bibnamefont {Pantelides}},\ }\href@noop {}
  {\bibfield  {journal} {\bibinfo  {journal} {IEEE Trans. Nucl. Sci.}\ }\textbf
  {\bibinfo {volume} {49}},\ \bibinfo {pages} {2650} (\bibinfo {year}
  {2002})}\BibitemShut {NoStop}%
\bibitem [{\citenamefont {Boch}\ \emph
  {et~al.}(2006{\natexlab{b}})\citenamefont {Boch}, \citenamefont {Saigne},
  \citenamefont {Touboul}, \citenamefont {Ducret}, \citenamefont {Carlotti},
  \citenamefont {Bernard}, \citenamefont {Schrimpf}, \citenamefont {Wrobel},\
  and\ \citenamefont {Sarrabayrouse}}]{boch2006dose}%
  \BibitemOpen
  \bibfield  {author} {\bibinfo {author} {\bibfnamefont {J.}~\bibnamefont
  {Boch}}, \bibinfo {author} {\bibfnamefont {F.}~\bibnamefont {Saigne}},
  \bibinfo {author} {\bibfnamefont {A.~D.}\ \bibnamefont {Touboul}}, \bibinfo
  {author} {\bibfnamefont {S.}~\bibnamefont {Ducret}}, \bibinfo {author}
  {\bibfnamefont {J.~F.}\ \bibnamefont {Carlotti}}, \bibinfo {author}
  {\bibfnamefont {M.}~\bibnamefont {Bernard}}, \bibinfo {author} {\bibfnamefont
  {R.~D.}\ \bibnamefont {Schrimpf}}, \bibinfo {author} {\bibfnamefont
  {F.}~\bibnamefont {Wrobel}}, \ and\ \bibinfo {author} {\bibfnamefont
  {G.}~\bibnamefont {Sarrabayrouse}},\ }\href@noop {} {\bibfield  {journal}
  {\bibinfo  {journal} {Appl. Phys. Lett.}\ }\textbf {\bibinfo {volume} {88}},\
  \bibinfo {pages} {232113} (\bibinfo {year} {2006}{\natexlab{b}})}\BibitemShut
  {NoStop}%
\bibitem [{\citenamefont {Bendler}\ and\ \citenamefont
  {Shlesinger}(1985)}]{bendler1985derivation}%
  \BibitemOpen
  \bibfield  {author} {\bibinfo {author} {\bibfnamefont {J.~T.}\ \bibnamefont
  {Bendler}}\ and\ \bibinfo {author} {\bibfnamefont {M.~F.}\ \bibnamefont
  {Shlesinger}},\ }\href@noop {} {\bibfield  {journal} {\bibinfo  {journal}
  {Macromolecules}\ }\textbf {\bibinfo {volume} {18}},\ \bibinfo {pages} {591}
  (\bibinfo {year} {1985})}\BibitemShut {NoStop}%
\bibitem [{\citenamefont {Scher}\ and\ \citenamefont
  {Montroll}(1975)}]{scher1975anomalous}%
  \BibitemOpen
  \bibfield  {author} {\bibinfo {author} {\bibfnamefont {H.}~\bibnamefont
  {Scher}}\ and\ \bibinfo {author} {\bibfnamefont {E.~W.}\ \bibnamefont
  {Montroll}},\ }\href@noop {} {\bibfield  {journal} {\bibinfo  {journal}
  {Phys. Rev. B}\ }\textbf {\bibinfo {volume} {12}},\ \bibinfo {pages} {2455}
  (\bibinfo {year} {1975})}\BibitemShut {NoStop}%
\bibitem [{\citenamefont {Shlesinger}\ and\ \citenamefont
  {Montroll}(1984)}]{shlesinger1984williams}%
  \BibitemOpen
  \bibfield  {author} {\bibinfo {author} {\bibfnamefont {M.~F.}\ \bibnamefont
  {Shlesinger}}\ and\ \bibinfo {author} {\bibfnamefont {E.~W.}\ \bibnamefont
  {Montroll}},\ }\href@noop {} {\bibfield  {journal} {\bibinfo  {journal}
  {Proc. Natl. Acad. Sci.}\ }\textbf {\bibinfo {volume} {81}},\ \bibinfo
  {pages} {1280} (\bibinfo {year} {1984})}\BibitemShut {NoStop}%
\bibitem [{\citenamefont {Pfister}\ and\ \citenamefont
  {Scher}(1978)}]{pfister1978dispersive}%
  \BibitemOpen
  \bibfield  {author} {\bibinfo {author} {\bibfnamefont {G.}~\bibnamefont
  {Pfister}}\ and\ \bibinfo {author} {\bibfnamefont {H.}~\bibnamefont
  {Scher}},\ }\href@noop {} {\bibfield  {journal} {\bibinfo  {journal} {Adv.
  Phys.}\ }\textbf {\bibinfo {volume} {27}},\ \bibinfo {pages} {747} (\bibinfo
  {year} {1978})}\BibitemShut {NoStop}%
\bibitem [{\citenamefont {Kakalios}\ \emph {et~al.}(1987)\citenamefont
  {Kakalios}, \citenamefont {Street},\ and\ \citenamefont
  {Jackson}}]{kakalios1987stretched}%
  \BibitemOpen
  \bibfield  {author} {\bibinfo {author} {\bibfnamefont {J.}~\bibnamefont
  {Kakalios}}, \bibinfo {author} {\bibfnamefont {R.}~\bibnamefont {Street}}, \
  and\ \bibinfo {author} {\bibfnamefont {W.}~\bibnamefont {Jackson}},\
  }\href@noop {} {\bibfield  {journal} {\bibinfo  {journal} {Phys. Rev. Lett.}\
  }\textbf {\bibinfo {volume} {59}},\ \bibinfo {pages} {1037} (\bibinfo {year}
  {1987})}\BibitemShut {NoStop}%
\bibitem [{\citenamefont {Fowler}(1956)}]{fowler1956x}%
  \BibitemOpen
  \bibfield  {author} {\bibinfo {author} {\bibfnamefont {J.~F.}\ \bibnamefont
  {Fowler}},\ }\href@noop {} {\bibfield  {journal} {\bibinfo  {journal}
  {Proceedings of the Royal Society of London. Series A. Mathematical and
  Physical Sciences}\ }\textbf {\bibinfo {volume} {236}},\ \bibinfo {pages}
  {464} (\bibinfo {year} {1956})}\BibitemShut {NoStop}%
\bibitem [{\citenamefont {Chen}\ \emph {et~al.}(1981)\citenamefont {Chen},
  \citenamefont {McKeever},\ and\ \citenamefont {Durrani}}]{chen1981solution}%
  \BibitemOpen
  \bibfield  {author} {\bibinfo {author} {\bibfnamefont {R.}~\bibnamefont
  {Chen}}, \bibinfo {author} {\bibfnamefont {S.}~\bibnamefont {McKeever}}, \
  and\ \bibinfo {author} {\bibfnamefont {S.}~\bibnamefont {Durrani}},\
  }\href@noop {} {\bibfield  {journal} {\bibinfo  {journal} {Phys. Rev. B}\
  }\textbf {\bibinfo {volume} {24}},\ \bibinfo {pages} {4931} (\bibinfo {year}
  {1981})}\BibitemShut {NoStop}%
\bibitem [{\citenamefont {Fowler}\ and\ \citenamefont
  {Farmer}(1954)}]{fowler1954conductivity-1}%
  \BibitemOpen
  \bibfield  {author} {\bibinfo {author} {\bibfnamefont {J.~F.}\ \bibnamefont
  {Fowler}}\ and\ \bibinfo {author} {\bibfnamefont {F.}~\bibnamefont
  {Farmer}},\ }\href@noop {} {\bibfield  {journal} {\bibinfo  {journal}
  {Nature}\ }\textbf {\bibinfo {volume} {173}},\ \bibinfo {pages} {317}
  (\bibinfo {year} {1954})}\BibitemShut {NoStop}%
\bibitem [{\citenamefont {Williams}\ and\ \citenamefont
  {Watts}(1970)}]{williams1970non}%
  \BibitemOpen
  \bibfield  {author} {\bibinfo {author} {\bibfnamefont {G.}~\bibnamefont
  {Williams}}\ and\ \bibinfo {author} {\bibfnamefont {D.~C.}\ \bibnamefont
  {Watts}},\ }\href@noop {} {\bibfield  {journal} {\bibinfo  {journal}
  {Transactions of the Faraday society}\ }\textbf {\bibinfo {volume} {66}},\
  \bibinfo {pages} {80} (\bibinfo {year} {1970})}\BibitemShut {NoStop}%
\bibitem [{\citenamefont {Kohlrausch}(1854)}]{kohlrausch1854theorie}%
  \BibitemOpen
  \bibfield  {author} {\bibinfo {author} {\bibfnamefont {R.}~\bibnamefont
  {Kohlrausch}},\ }\href@noop {} {\bibfield  {journal} {\bibinfo  {journal}
  {Ann. Phys.}\ }\textbf {\bibinfo {volume} {167}},\ \bibinfo {pages} {179}
  (\bibinfo {year} {1854})}\BibitemShut {NoStop}%
\bibitem [{\citenamefont {Griscom}(1990)}]{griscom1990growth}%
  \BibitemOpen
  \bibfield  {author} {\bibinfo {author} {\bibfnamefont {D.~L.}\ \bibnamefont
  {Griscom}},\ }\href@noop {} {\bibfield  {journal} {\bibinfo  {journal} {Nucl.
  Instrum. Methods Phys. Res. Sect. B Beam Interact. Mater. At.}\ }\textbf
  {\bibinfo {volume} {46}},\ \bibinfo {pages} {12} (\bibinfo {year}
  {1990})}\BibitemShut {NoStop}%
\bibitem [{\citenamefont {Imai}\ and\ \citenamefont
  {Hirashima}(1994)}]{imai1994intrinsic}%
  \BibitemOpen
  \bibfield  {author} {\bibinfo {author} {\bibfnamefont {H.}~\bibnamefont
  {Imai}}\ and\ \bibinfo {author} {\bibfnamefont {H.}~\bibnamefont
  {Hirashima}},\ }\href@noop {} {\bibfield  {journal} {\bibinfo  {journal} {J.
  Non. Cryst. Solids}\ }\textbf {\bibinfo {volume} {179}},\ \bibinfo {pages}
  {202} (\bibinfo {year} {1994})}\BibitemShut {NoStop}%
\bibitem [{\citenamefont {Mashkov}\ \emph {et~al.}(1996)\citenamefont
  {Mashkov}, \citenamefont {Austin}, \citenamefont {Zhang},\ and\ \citenamefont
  {Leisure}}]{mashkov1996fundamental}%
  \BibitemOpen
  \bibfield  {author} {\bibinfo {author} {\bibfnamefont {V.}~\bibnamefont
  {Mashkov}}, \bibinfo {author} {\bibfnamefont {W.~R.}\ \bibnamefont {Austin}},
  \bibinfo {author} {\bibfnamefont {L.}~\bibnamefont {Zhang}}, \ and\ \bibinfo
  {author} {\bibfnamefont {R.}~\bibnamefont {Leisure}},\ }\href@noop {}
  {\bibfield  {journal} {\bibinfo  {journal} {Phys. Rev. Lett.}\ }\textbf
  {\bibinfo {volume} {76}},\ \bibinfo {pages} {2926} (\bibinfo {year}
  {1996})}\BibitemShut {NoStop}%
\bibitem [{\citenamefont {Kimerling}(1978)}]{kimerling1978recombination}%
  \BibitemOpen
  \bibfield  {author} {\bibinfo {author} {\bibfnamefont {L.}~\bibnamefont
  {Kimerling}},\ }\href@noop {} {\bibfield  {journal} {\bibinfo  {journal}
  {Solid-State Electron.}\ }\textbf {\bibinfo {volume} {21}},\ \bibinfo {pages}
  {1391} (\bibinfo {year} {1978})}\BibitemShut {NoStop}%
\bibitem [{\citenamefont {Weeks}\ \emph {et~al.}(1975)\citenamefont {Weeks},
  \citenamefont {Tully},\ and\ \citenamefont {Kimerling}}]{weeks1975theory}%
  \BibitemOpen
  \bibfield  {author} {\bibinfo {author} {\bibfnamefont {J.~D.}\ \bibnamefont
  {Weeks}}, \bibinfo {author} {\bibfnamefont {J.~C.}\ \bibnamefont {Tully}}, \
  and\ \bibinfo {author} {\bibfnamefont {L.}~\bibnamefont {Kimerling}},\
  }\href@noop {} {\bibfield  {journal} {\bibinfo  {journal} {Phys. Rev. B}\
  }\textbf {\bibinfo {volume} {12}},\ \bibinfo {pages} {3286} (\bibinfo {year}
  {1975})}\BibitemShut {NoStop}%
\bibitem [{\citenamefont {Lang}\ and\ \citenamefont
  {Kimerling}(1974)}]{lang1974observation}%
  \BibitemOpen
  \bibfield  {author} {\bibinfo {author} {\bibfnamefont {D.}~\bibnamefont
  {Lang}}\ and\ \bibinfo {author} {\bibfnamefont {L.}~\bibnamefont
  {Kimerling}},\ }\href@noop {} {\bibfield  {journal} {\bibinfo  {journal}
  {Phys. Rev. Lett.}\ }\textbf {\bibinfo {volume} {33}},\ \bibinfo {pages}
  {489} (\bibinfo {year} {1974})}\BibitemShut {NoStop}%
\bibitem [{\citenamefont {Zhang}\ \emph {et~al.}(1995)\citenamefont {Zhang},
  \citenamefont {Mashkov},\ and\ \citenamefont {Leisure}}]{zhang1995multiple}%
  \BibitemOpen
  \bibfield  {author} {\bibinfo {author} {\bibfnamefont {L.}~\bibnamefont
  {Zhang}}, \bibinfo {author} {\bibfnamefont {V.}~\bibnamefont {Mashkov}}, \
  and\ \bibinfo {author} {\bibfnamefont {R.}~\bibnamefont {Leisure}},\
  }\href@noop {} {\bibfield  {journal} {\bibinfo  {journal} {Phys. Rev. Lett.}\
  }\textbf {\bibinfo {volume} {74}},\ \bibinfo {pages} {1605} (\bibinfo {year}
  {1995})}\BibitemShut {NoStop}%
\bibitem [{\citenamefont {Pease}\ \emph {et~al.}(2004)\citenamefont {Pease},
  \citenamefont {Platteter}, \citenamefont {Dunham}, \citenamefont {Seiler},
  \citenamefont {Barnaby}, \citenamefont {Schrimpf}, \citenamefont
  {Shaneyfelt}, \citenamefont {Maher},\ and\ \citenamefont
  {Nowlin}}]{pease2004characterization}%
  \BibitemOpen
  \bibfield  {author} {\bibinfo {author} {\bibfnamefont {R.~L.}\ \bibnamefont
  {Pease}}, \bibinfo {author} {\bibfnamefont {D.~G.}\ \bibnamefont
  {Platteter}}, \bibinfo {author} {\bibfnamefont {G.}~\bibnamefont {Dunham}},
  \bibinfo {author} {\bibfnamefont {J.}~\bibnamefont {Seiler}}, \bibinfo
  {author} {\bibfnamefont {H.}~\bibnamefont {Barnaby}}, \bibinfo {author}
  {\bibfnamefont {R.}~\bibnamefont {Schrimpf}}, \bibinfo {author}
  {\bibfnamefont {M.~R.}\ \bibnamefont {Shaneyfelt}}, \bibinfo {author}
  {\bibfnamefont {M.}~\bibnamefont {Maher}}, \ and\ \bibinfo {author}
  {\bibfnamefont {R.}~\bibnamefont {Nowlin}},\ }\href@noop {} {\bibfield
  {journal} {\bibinfo  {journal} {IEEE Trans. Nucl. Sci.}\ }\textbf {\bibinfo
  {volume} {51}},\ \bibinfo {pages} {3773} (\bibinfo {year}
  {2004})}\BibitemShut {NoStop}%
\bibitem [{\citenamefont {McWhorter}\ and\ \citenamefont
  {Winokur}(1986)}]{mcwhorter1986simple}%
  \BibitemOpen
  \bibfield  {author} {\bibinfo {author} {\bibfnamefont {P.}~\bibnamefont
  {McWhorter}}\ and\ \bibinfo {author} {\bibfnamefont {P.}~\bibnamefont
  {Winokur}},\ }\href@noop {} {\bibfield  {journal} {\bibinfo  {journal} {Appl.
  Phys. Lett.}\ }\textbf {\bibinfo {volume} {48}},\ \bibinfo {pages} {133}
  (\bibinfo {year} {1986})}\BibitemShut {NoStop}%
\bibitem [{\citenamefont {Ball}\ \emph {et~al.}(2002)\citenamefont {Ball},
  \citenamefont {Schrimpf},\ and\ \citenamefont
  {Barnaby}}]{ball2002separation}%
  \BibitemOpen
  \bibfield  {author} {\bibinfo {author} {\bibfnamefont {D.~R.}\ \bibnamefont
  {Ball}}, \bibinfo {author} {\bibfnamefont {R.~D.}\ \bibnamefont {Schrimpf}},
  \ and\ \bibinfo {author} {\bibfnamefont {H.~J.}\ \bibnamefont {Barnaby}},\
  }\href@noop {} {\bibfield  {journal} {\bibinfo  {journal} {IEEE Trans. Nucl.
  Sci.}\ }\textbf {\bibinfo {volume} {49}},\ \bibinfo {pages} {3185} (\bibinfo
  {year} {2002})}\BibitemShut {NoStop}%
\bibitem [{\citenamefont {Song}\ and\ \citenamefont
  {Wei}(2020)}]{song2020origin}%
  \BibitemOpen
  \bibfield  {author} {\bibinfo {author} {\bibfnamefont {Y.}~\bibnamefont
  {Song}}\ and\ \bibinfo {author} {\bibfnamefont {S.-H.}\ \bibnamefont {Wei}},\
  }\href@noop {} {\bibfield  {journal} {\bibinfo  {journal} {ACS Appl.
  Electron. Mater.}\ }\textbf {\bibinfo {volume} {2}},\ \bibinfo {pages} {3783}
  (\bibinfo {year} {2020})}\BibitemShut {NoStop}%
\bibitem [{\citenamefont {Song}\ \emph {et~al.}(2020)\citenamefont {Song},
  \citenamefont {Zhou}, \citenamefont {Cai}, \citenamefont {Liu}, \citenamefont
  {Yang}, \citenamefont {Zhang}, \citenamefont {Zhang}, \citenamefont {Lan},\
  and\ \citenamefont {Wei}}]{song2020defect}%
  \BibitemOpen
  \bibfield  {author} {\bibinfo {author} {\bibfnamefont {Y.}~\bibnamefont
  {Song}}, \bibinfo {author} {\bibfnamefont {H.}~\bibnamefont {Zhou}}, \bibinfo
  {author} {\bibfnamefont {X.-F.}\ \bibnamefont {Cai}}, \bibinfo {author}
  {\bibfnamefont {Y.}~\bibnamefont {Liu}}, \bibinfo {author} {\bibfnamefont
  {P.}~\bibnamefont {Yang}}, \bibinfo {author} {\bibfnamefont {G.-H.}\
  \bibnamefont {Zhang}}, \bibinfo {author} {\bibfnamefont {Y.}~\bibnamefont
  {Zhang}}, \bibinfo {author} {\bibfnamefont {M.}~\bibnamefont {Lan}}, \ and\
  \bibinfo {author} {\bibfnamefont {S.-H.}\ \bibnamefont {Wei}},\ }\href@noop
  {} {\bibfield  {journal} {\bibinfo  {journal} {ACS Appl. Mater. Interfaces}\
  }\textbf {\bibinfo {volume} {12}},\ \bibinfo {pages} {29993} (\bibinfo {year}
  {2020})}\BibitemShut {NoStop}%
\bibitem [{\citenamefont {Winokur}\ and\ \citenamefont
  {Boesch}(1980)}]{winokur1980interface}%
  \BibitemOpen
  \bibfield  {author} {\bibinfo {author} {\bibfnamefont {P.}~\bibnamefont
  {Winokur}}\ and\ \bibinfo {author} {\bibfnamefont {H.}~\bibnamefont
  {Boesch}},\ }\href@noop {} {\bibfield  {journal} {\bibinfo  {journal} {IEEE
  Trans. Nucl. Sci.}\ }\textbf {\bibinfo {volume} {27}},\ \bibinfo {pages}
  {1647} (\bibinfo {year} {1980})}\BibitemShut {NoStop}%
\end{thebibliography}

%

\section{supplementary materials}

\begin{figure}[!h]
  \centering
  \includegraphics[width=0.87\linewidth]{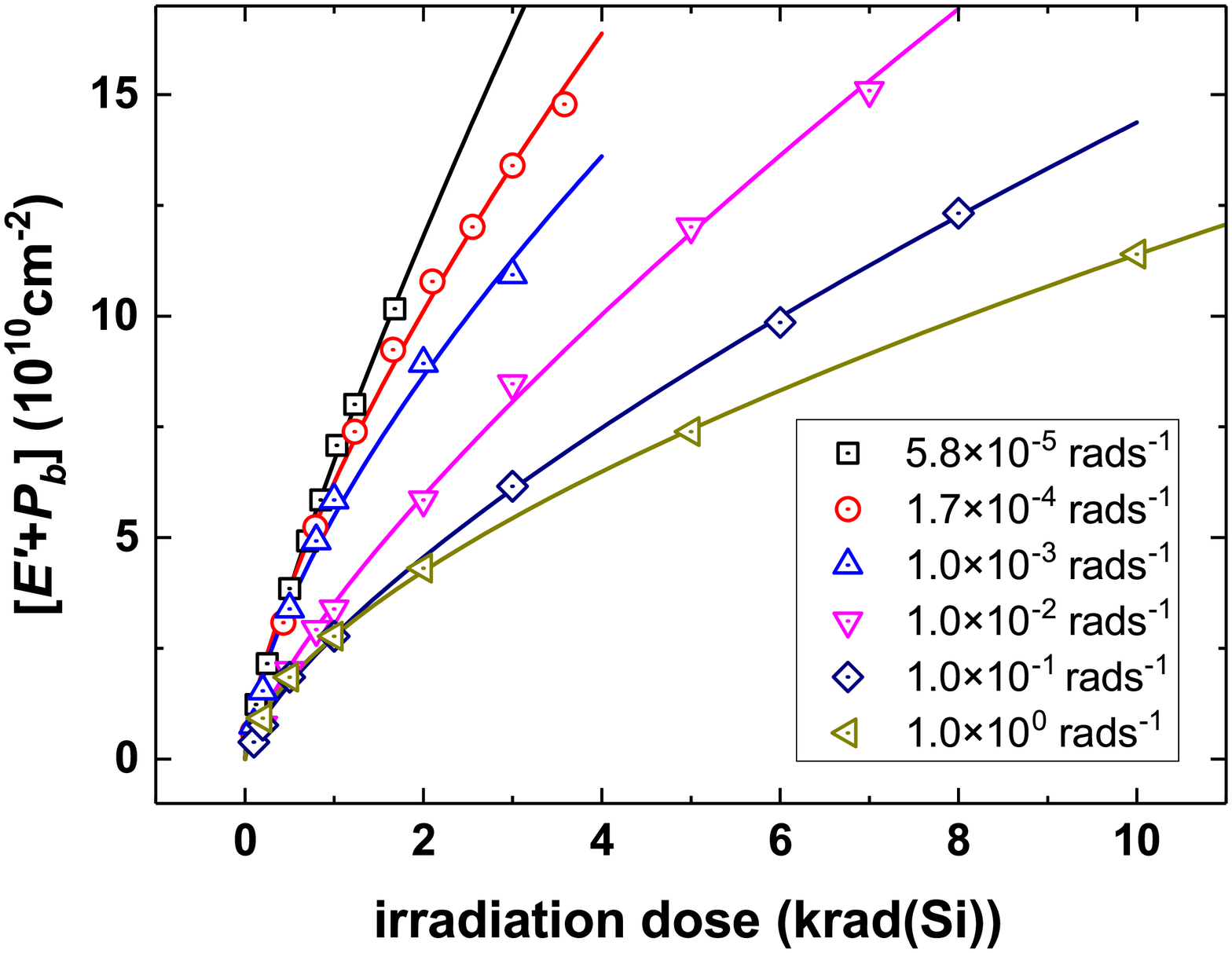}\\
\begin{flushleft}
Fig. S1. The total concentration of $E'$ and $P_b$ centers as a function of the irradiation dose at different dose rates.  Dots are for data and curves are for fitting.
\end{flushleft}
\end{figure}

\begin{figure}[!h]
  \centering
  \includegraphics[width=0.9\linewidth]{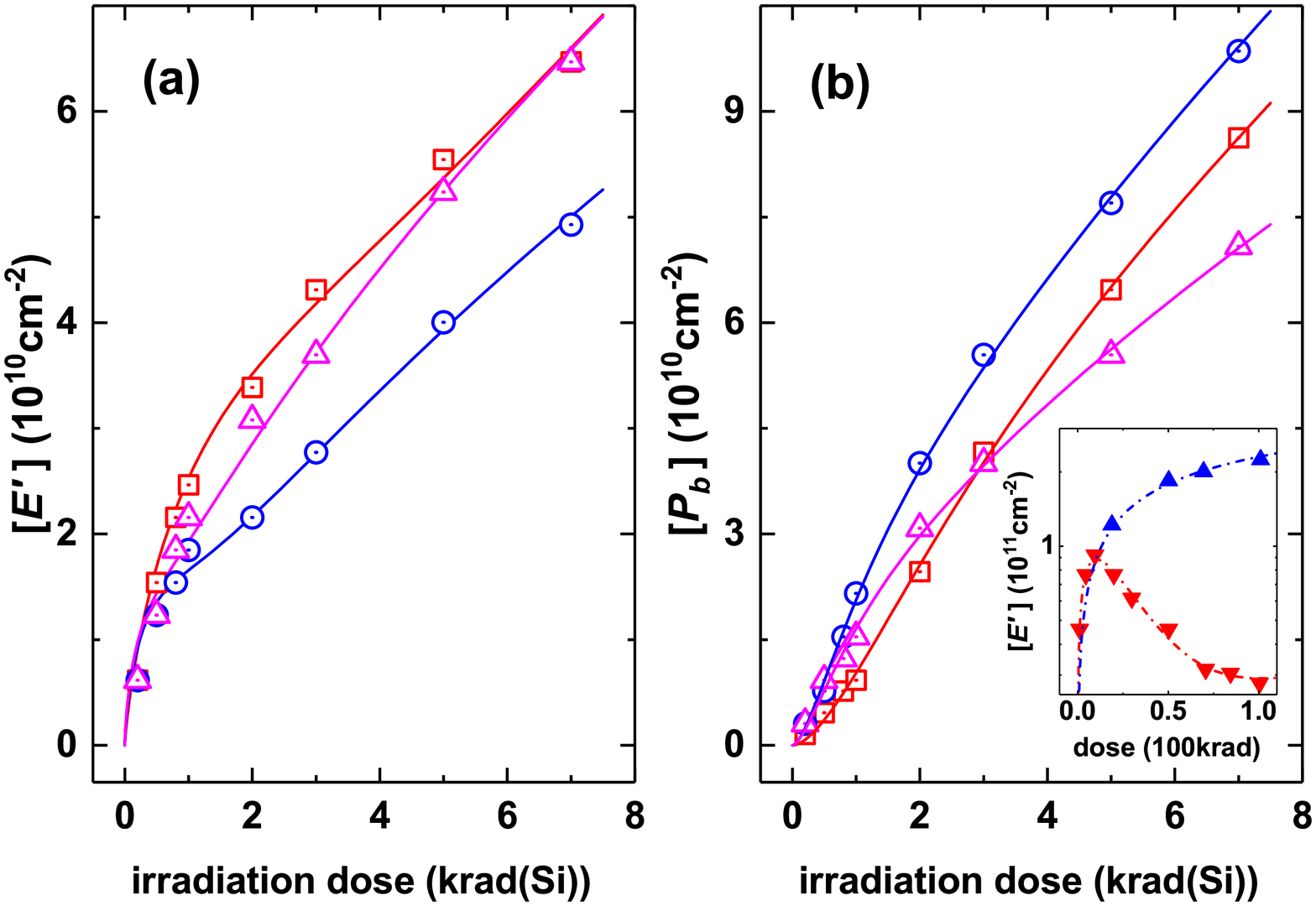}\\ 
\begin{flushleft}
Fig. S2. The concentration of $E'$ center (a) and $P_b$ center (b) as a function of ionizing dose at 10 mrad/s for three different samples.
  Insert in (b): the fitting of Li et al's $E'$ data~[33].
  Dots are for data and curves are for fitting.
\end{flushleft}
\end{figure}

\begin{figure}[!h]
  \centering
  \includegraphics[width=0.9\linewidth]{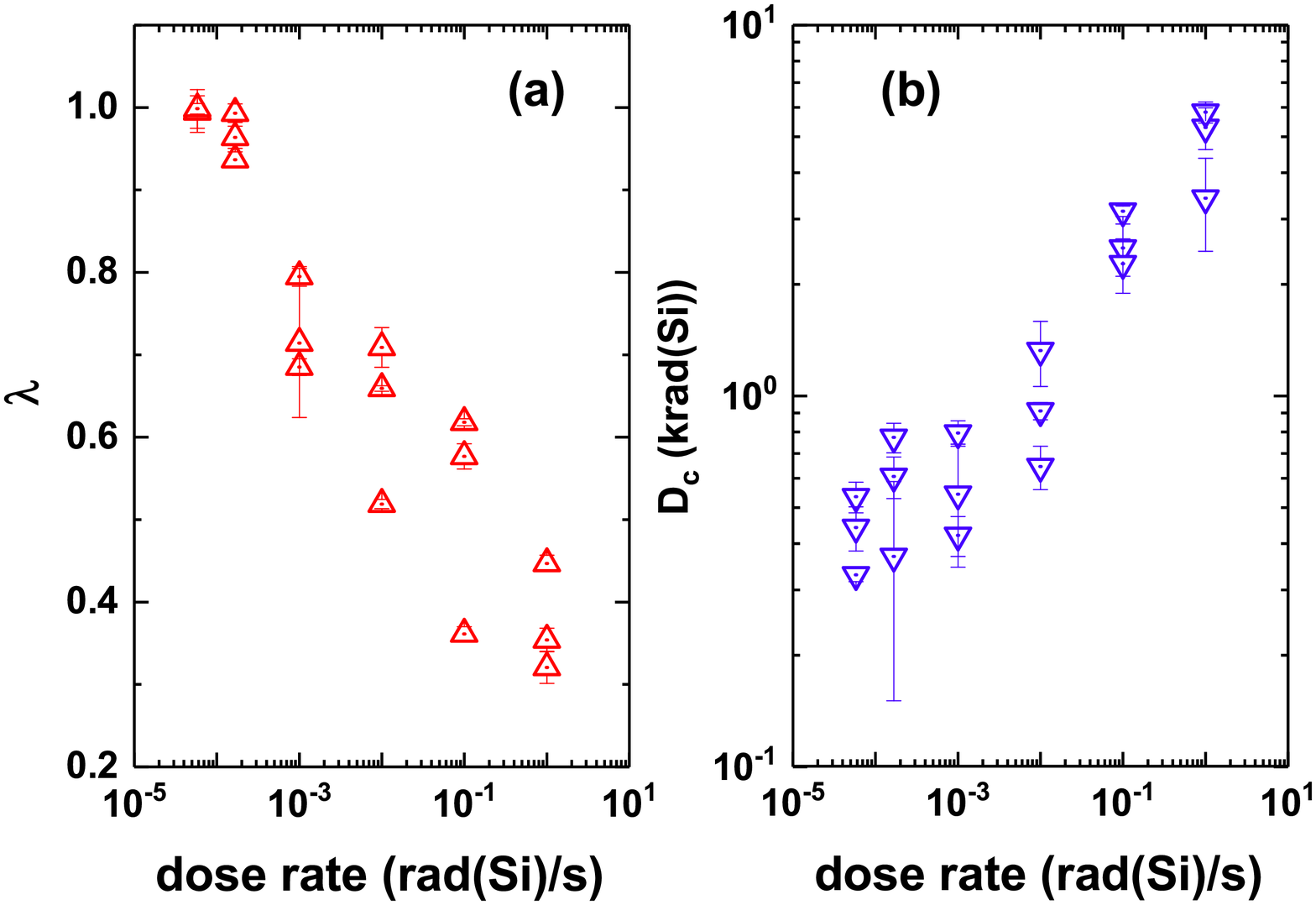}\\
\begin{flushleft}
Fig. S3. The conversion ratio (a) and characteristic conversion dose (b) extracted from the data.
\end{flushleft}
\end{figure}

\end{document}